\documentclass[
    reprint,
	amsmath,
    amssymb,
	aps,
	prx,
    floatfix,
    superscriptaddress
]{revtex4-2}

\usepackage[english]{babel}
\usepackage[utf8x]{inputenc}
\usepackage[T1]{fontenc}
\usepackage{braket}
\usepackage{mathtools}
\usepackage{flushend}

\usepackage{amsmath}
\usepackage{siunitx}
\usepackage{graphicx}
\usepackage{dcolumn}
\usepackage{bm}
\usepackage{array}
\usepackage[shortlabels]{enumitem}
\newcolumntype{?}{!{\vrule width 1pt}}
\usepackage[colorinlistoftodos]{todonotes}
\usepackage[colorlinks=true,linkcolor=blue,citecolor=blue,urlcolor=blue]{hyperref}
\usepackage{slashed}

\makeatletter
\newcommand*{\newbibstartnumber}[1]{%
  \apptocmd{\thebibliography}{%
    \global\c@NAT@ctr #1\relax
    \addtocounter{NAT@ctr}{-1}%
  }{}{}%
}
\makeatother

\makeatletter
\let\cat@comma@active\@empty
\makeatother
\begin{document}
\title{Topological Signatures in Nodal Semimetals through Neutron Scattering}
\author{Thanh Nguyen}
\thanks{These authors contributed equally to this work.}
\affiliation{Department of Nuclear Science and Engineering, MIT, Cambridge, MA 02139, USA}
\author{Yoichiro Tsurimaki}
\thanks{These authors contributed equally to this work.}
\affiliation{Department of Mechanical Engineering, MIT, Cambridge, MA 02139, USA}
\affiliation{Department of Physics, Stanford University, Stanford, CA 94305, USA}
\author{Ricardo Pablo-Pedro}
\thanks{These authors contributed equally to this work.}
\affiliation{Department of Nuclear Science and Engineering, MIT, Cambridge, MA 02139, USA}
\author{Grigory Bednik}
\thanks{These authors contributed equally to this work.}
\affiliation{Department of Physics, University of California, Santa Cruz, Santa Cruz, CA 95064, USA}
\author{Tongtong Liu}
\affiliation{Department of Physics, MIT, Cambridge, MA 02139, USA}
\author{Anuj Apte}
\altaffiliation{Present address: Department of Physics, University of Chicago, Chicago, IL 60637, USA}
\affiliation{Department of Physics, MIT, Cambridge, MA 02139, USA}
\author{Nina Andrejevic}
\affiliation{Department of Material Science and Engineering, MIT, Cambridge, MA 02139, USA}
\author{Mingda Li}
\thanks{Corresponding author.\\\href{mailto:mingda@mit.edu}{mingda@mit.edu} \vspace{0.5cm}}
\affiliation{Department of Nuclear Science and Engineering, MIT, Cambridge, MA 02139, USA}
\date{\today}

\begin{abstract}
Topological nodal semimetals are known to host a variety of fascinating electronic properties due to the topological protection of the band-touching nodes. Neutron scattering, despite its power in probing elementary excitations, has not been routinely applied to topological semimetals, mainly due to the lack of an explicit connection between the neutron response and the signature of topology. In this work, we theoretically investigate the role that neutron scattering can play to unveil the topological nodal features: a large magnetic neutron response with spectral non-analyticity can be generated solely from the nodal bands. A new formula for the dynamical structure factor for generic topological nodal metals is derived. For Weyl semimetals, we show that the locations of Weyl nodes, the Fermi velocities and the signature of chiral anomaly can all leave hallmark neutron spectral responses. Our work offers a neutron-based avenue towards probing bulk topological materials.
\end{abstract}

\maketitle


\section{Introduction}
Topological materials encompass broad categories of materials where nontrivial topology plays a fundamental role in electronic and functional properties. In three-dimensional (3D) materials, insulating topological materials include topological insulators (TIs) \cite{fu2007,chen2009,hsieh2009,kitaev2009,hasan2010,moore2010,qi2011,bernevig2013,ando2013,bansil2016,schindler2018} and topological crystalline insulators (TCIs) \cite{fu2011,dziawa2012,hsieh2012,xu2012,tanaka2012}; while for metals and semimetals, topological materials contain gapless nodes, such as in nodal-point Dirac (DSMs) \cite{wang2012,wang2013,neupane2014,borisenko2014,liu2014,liu2014-1}, Weyl semimetals (WSMs) \cite{lv2015,lv2015-1,xu2015,xu2015-1,lu2015,weng2015,huang2015,armitage2018} and in nodal-line semimetals (NLSMs) \cite{burkov2011,fang2015,fang2016,bian2016,hu2016,neupane2016,schoop2016}, among numerous topological nodal metals (TNMs) hosting a variety of emergent unconventional quasiparticles \cite{bradlyn2016,bzdusek2016,wang2016,lv2017}.

For some time, the discovery of new topological materials was largely conducted on a case-by-case basis, as was the case for the prototypical Bi$_2$Se$_3$ TI family (with members Sb$_2$Te$_3$ and Bi$_2$Te$_3$) and the TaAs WSM family (with members TaP, NbP, and NbAs). Recent theoretical progress in symmetry-based indicators \cite{kruthoff2017,po2017,watanabe2018} and topological quantum chemistry \cite{bradlyn2017} has profoundly accelerated the theoretical search of nontrivial topology within materials. Through symmetry-based indicators, bandstructures are interpreted in terms of elementary basis states whereby the topologically-trivial atomic insulator states are removed allowing the nontrivial topological bands to be revealed. On the other hand, topological quantum chemistry provides a holistic description of the global properties for all possible bandstructures, using a graph-theoretical description of momentum space and a dual group one for real space. Using these approaches, recent high-throughput calculations have obtained a largely-broad family of materials that may carry nontrivial topology in both non-magnetic \cite{zhang2019,vergniory2019,tang2019} and, more recently, magnetic materials \cite{xu2020}. 

The fact that approximately 30\% of all 3D stoichiometric materials may carry nontrivial topology brings us to the question of experimental verification. In contrast to the extensive developments on the theoretical forefront, the experimental determination of nontrivial topology remains an arduous challenge. Angular-resolved photoemission spectroscopy (ARPES) plays a significant role in probing topology by directly measuring the bandstructures and afterward, comparing with \textit{ab initio} calculations, but a few difficulties exist \cite{damascelli2003,suga2014,lv2019}. Fundamentally, only materials that can cleave without difficulty can be probed by ARPES. This limits the types of materials that can be measured with this technique, since many materials do not form the required atomically-flat and well-ordered surface following cleavage. On top of this, the signals from photoelectrons prohibit the application of a magnetic field to study field-driven phenomena. From a practical viewpoint, the high vacuum and the atomically flat surface pose high technical barriers for sample preparation. Other experimental probes, such as scanning tunneling microscopy (STM), pose an even higher technical impediment due to the necessity of protecting against minuscule vibrations, while quantum transport, despite being routinely carried out, is generally considered as an indirect method to infer bulk topology in TNMs \cite{murakawa2013,zhao2015}. Given the large number of topological material candidates, an alternative experimental approach that can address the aforementioned challenges of existing techniques, but still capable of capturing the non-trivial topology, would be highly desirable.

\begin{figure}[h!]
	\centering
	\includegraphics[width=\linewidth]{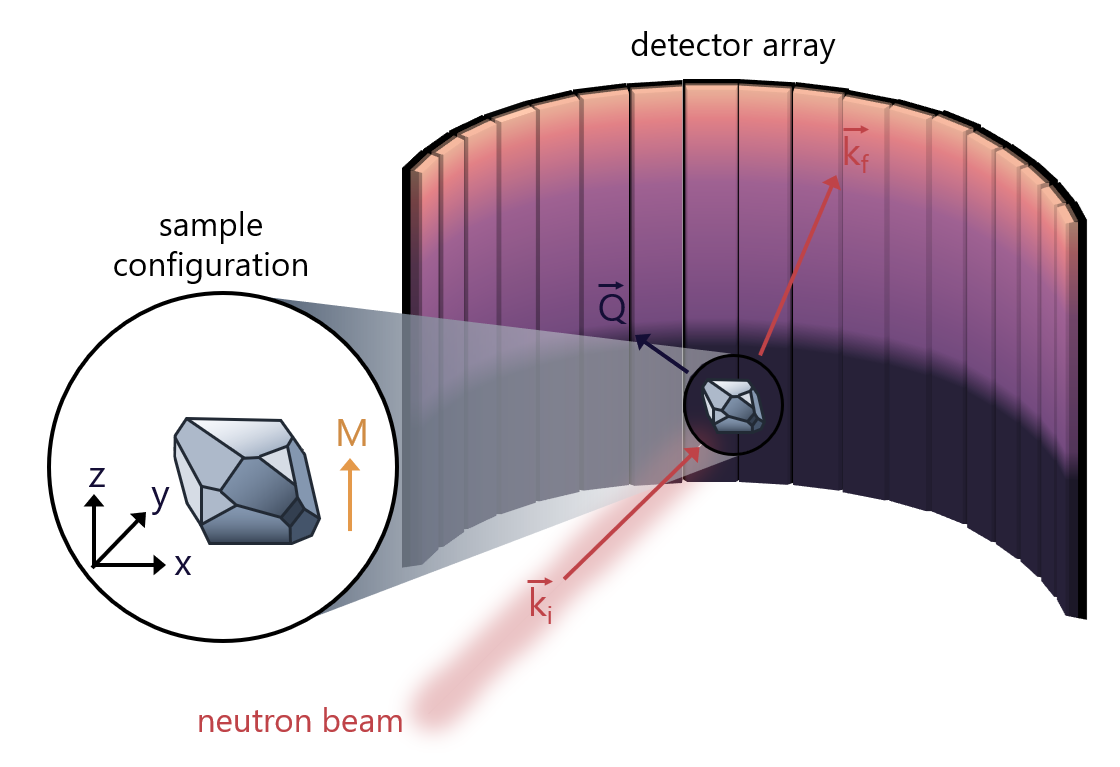}
	\caption{\textbf{Experimental configuration.} Nodal signatures of topology can be observed in the neutron spectra, e.g., extended non-analytical lines of intensity, in the case of a time-of-flight neutron experiment schematically shown here.}
	\label{fig:1}
\end{figure}

In this work, we demonstrate that neutron scattering is a powerful, yet unexplored, tool to capture the topological nature of TNMs (Fig. \ref{fig:1}). Through quantum single-body calculations, we derive a new dynamical structure factor formula for generic TNMs, showing that they can have a large magnetic neutron scattering response. This solely arises from the nodal bands which potentially carry nontrivial topology. This contrasts with the conventional Fermi liquid in which the magnetic signal from itinerant electrons is negligibly small \cite{lovesey1986}. In particular, a closed-form expression for the dynamical structure factor is obtained for prototypical WSMs, whereby we show that the Weyl node locations in the Brillouin zone and the magnitude of the Fermi velocities will leave hallmark signatures in neutron spectra through their momentum space separation and their chemical potential. Moreover, when both electrical and magnetic fields are applied, we show that the chiral anomaly can leave measurable fingerprints on the spectra. These features retain for a realistic WSM containing multiple Weyl nodes and are generalizable to other types of TNMs which exhibit nodal lines or surfaces. Given that neutron scattering can directly measure bulk samples and is compatible with external fields, our proposed work offers a new avenue towards measuring TNMs with advantages such as the ability to probe a broader repertoire of TNMs which are only sensitive to topological bands, and to study possible field-driven topological phase transitions in these exotic materials.

The rest of the paper is organized as follows. In Section \ref{sec:generic}, we begin by deriving a formula of the dynamical structure factor for a generic TNM and subsequently, in Section \ref{sec:prototypical}, we expand on this to compute this quantity for a toy model of Weyl fermions. A model in the case of a nodal line semimetal is provided in Section \ref{sec:nodalline}. We then calculate the corresponding neutron response function for realistic WSMs with multiple pairs of Weyl nodes in the Brillouin zone through which the neutron response can reveal their location (Section \ref{sec:locations}) and offer an approach to experimentally observe a signature of chiral anomaly (Section \ref{sec:chiralanomaly}). Effects of finite mass in DSMs and finite temperature on the neutron response are investigated in Sections \ref{sec:finitemass} and \ref{sec:finiteT}, respectively. We holistically discuss our results in Section \ref{sec:discussion} on the large magnetic neutron scattering response resulting from nontrivial topology in TNMs with a WSM example and possible candidate materials. We conclude in Section \ref{sec:conclusion} with a general discussion on how our results fit in with current experimental techniques used to probe TNMs. Appendices contain further details of calculations.

\section{Dynamical Structure Factor for Generic Topological Nodal Metals}\label{sec:generic} We start with a general linear-quadratic Hamiltonian which can be used to describe TNMs
\begin{equation}
\begin{split}
    \label{eq:genericH}
    H =\ &\int d^3\boldsymbol{r}\ \sum_{\alpha\beta}\psi_\alpha^\dagger(\boldsymbol{r})\Big[\sum_iv_i^{\alpha\beta}\big(-i\hbar\partial_i-a_i^{\alpha\beta}\big)+\\
    & \sum_{ij}b_{ij}^{\alpha\beta}\left(-i\hbar\partial_i\right)\left(-i\hbar\partial_j\right)\Big]\psi_\beta(\boldsymbol{r})
\end{split}
\end{equation}
\noindent where $\psi^\dagger(\boldsymbol{r})$ ($\psi(\boldsymbol{r})$) creates (annihilates) an electron at position $\boldsymbol{r}$; $\alpha$, $\beta$ denote general spin and band indices; $i$, $j=1$, $2$, $3$ label spatial directions; $\hbar$ is the reduced Planck constant; $v_i^{\alpha\beta}$ is the nodal term of the generalized Fermi velocities; $a_i^{\alpha\beta}$ is the node shift; and $b_{ij}^{\alpha\beta}$ is a conventional quadratic term.

To describe the influence of a magnetic field, we adopt the minimal coupling prescription $i\hbar\partial_i\rightarrow i\hbar\partial_i + eA_i$ in Eq. \eqref{eq:genericH}. Within the linear-response regime, the current operator in momentum space can be written as
\begin{equation}
    \label{eq:currentop}
    J_i(\boldsymbol{Q})= -\frac{e}{V}\sum_{\boldsymbol{k}\alpha\beta}\psi_{\boldsymbol{k}\alpha}^\dagger\Big[v_i^{\alpha\beta}+\sum_j \hbar b_{ij}^{\alpha\beta}\left(2 k_j- Q_j\right)\Big]\psi_{\boldsymbol{k}-\boldsymbol{Q}\beta}
\end{equation}
\noindent where $V$ is the volume, $\boldsymbol{k}$ is the wavevector and $\boldsymbol{Q}\equiv\boldsymbol{k}_i-\boldsymbol{k}_f$ is the momentum transfer between initial and final states which is also the neutron momentum transfer. 

Following the derivation sketched in Appendix \ref{ap:A}, the dimensionless magnetization obtained from the current operator in the case of a TNM can be written as
\begin{equation}
    \label{eq:magnetization}
    \boldsymbol{M}(\boldsymbol{Q}) = -\frac{i}{2\mu_B c}\frac{\boldsymbol{J}(\boldsymbol{Q})\times\boldsymbol{Q}}{Q^2}
\end{equation}
\noindent where $\mu_B$ is the Bohr magneton in Gaussian units and $Q=|\boldsymbol{Q}|$. The corresponding magnetic dynamical structure factor tensor can be written into the following form
\begin{equation}
    \label{eq:magneticdynamicstructurefactor}
    S_{jl}(\boldsymbol{Q},\omega) = \frac{1}{2\pi}\int_{-\infty}^{\infty}dt\ e^{-i\omega t}\left\langle M_j(-\boldsymbol{Q},0)M_l(\boldsymbol{Q},t) \right\rangle
\end{equation}
\noindent where $\omega$ is the frequency of the neutron and $M_l(\boldsymbol{Q},t)=e^{iHt/\hbar}M_l(\boldsymbol{Q},0)e^{-iHt/\hbar}$ is the neutron magnetization operator in the Heisenberg picture. By comparison, the magnetization operator comprising of localized moments and itinerant electrons is written as \cite{lovesey1984}
\begin{equation}
    \label{eq:magnetizationlocalized}
    \boldsymbol{M}(\boldsymbol{Q})=\sum_Ne^{i\boldsymbol{Q}\cdot\boldsymbol{r}_N}\left(\boldsymbol{s}_N + i\frac{\boldsymbol{k}_{e,N}\times\boldsymbol{Q}}{Q^2}\right)
\end{equation}
\noindent where the sum runs over all local sites $N$ and $\boldsymbol{s}_N$, $\boldsymbol{k}_{e,N}$ are the local spin and electron wavevector, respectively. For non-relativistic electrons, we have $\boldsymbol{J}=-e\hbar\boldsymbol{k}/m_e$ where $m_e$ is the electron mass. Our result in Eq. \eqref{eq:magnetization} for TNMs thus resembles the magnetization for itinerant electrons in Eq. \eqref{eq:magnetizationlocalized}.

However, there is one major difference between Eqs. \eqref{eq:magnetization} and \eqref{eq:magnetizationlocalized} that enables TNMs to induce a large neutron scattering signal. In conventional Fermi liquids (second term of Eq. \eqref{eq:magnetizationlocalized}), the total magnetization that arises from different itinerant electrons $\boldsymbol{k}_{e,N}$ cancels out, leading to a small neutron magnetic scattering signal \cite{lovesey1986}. To understand this, taking the $Q \to 0$ limit where magnetic signal is large, the magnetization is $\boldsymbol{M} \sim i\sum_N \boldsymbol{k}_{e,N}/Q$, and the magnetic signal from an electron with momentum $\boldsymbol{k}_{e,N}$ cancels out the $-\boldsymbol{k}_{e,N}$ contribution from another electron. In contrast, for TNMs, topology plays a role to protect the gapless nodal features, in which different charge carriers can contribute to the magnetization in an additive manner ($v_i^{\alpha\beta}$ term in Eq. \eqref{eq:currentop}). This leads to a large non-cancellation effect of the corresponding magnetization and thus, to a large neutron signal. For instance, in a simple WSM, the corresponding $v_i^{\alpha\beta}$ term that generates the current is simply the Fermi velocity of the Weyl fermions. 

To facilitate the calculations, we rewrite the dynamical structure factor tensor of Eq. \eqref{eq:magneticdynamicstructurefactor} as
\begin{equation}
\begin{split}
    \label{eq:mdsf}
    S_{jl}(\boldsymbol{Q},&\omega)=\ -\frac{\hbar V}{\pi(2\mu_B c)^2}\frac{1}{1-e^{-\beta\hbar\omega}}\times\\
    &\sum_{ikmn}\epsilon_{jik}\epsilon_{lmn}\frac{Q_kQ_n}{Q^4}\text{Im}\left[\Pi_{im}(\boldsymbol{Q},\omega)\right]
\end{split}
\end{equation}
\noindent where $\beta=(k_BT)^{-1}$, $k_B$ is the Boltzmann constant, $T$ is the temperature, $\epsilon_{jik}$ is the Levi-Civita symbol, and $\Pi_{im}(\boldsymbol{Q},\omega)$ is the dynamical response function (also known as the magnetic susceptibility tensor or the current-current correlation function) defined as
\begin{equation}
    \label{eq:magneticsusceptibility}
    \left.\Pi_{im}(\boldsymbol{Q},\omega)=\int_0^\beta d\tau\ e^{-i\omega_n\tau}\Pi_{im}(\boldsymbol{Q},\tau)\right|_{i\omega_n\rightarrow\hbar\omega+i0^{+}}
\end{equation}
\noindent where $\omega_n=2\pi n/\beta$ ($n=0$, $\pm1$, $\pm2$, $\dots$) is the bosonic Matsubara frequency. Eq. \eqref{eq:magneticsusceptibility} originates from the imaginary-time current-current correlation function
\begin{equation}
    \label{eq:current-current}
    \Pi_{im}(\boldsymbol{Q},\tau)=-\frac{1}{V}\left\langle T_{\tau}J_i(-\boldsymbol{Q}, 0)J_m(\boldsymbol{Q},\tau)\right\rangle
\end{equation}
\noindent where $T_\tau$ is the imaginary time-ordering operator, $\left\langle\dots\right\rangle$ denotes the imaginary time average over the whole canonical ensemble, and the total current operator is given in the Heisenberg picture as $J_m(\boldsymbol{Q},\tau) = e^{H\tau}J_m(\boldsymbol{Q})e^{-H\tau}$. Further details of the calculation of Eq. \eqref{eq:mdsf} can be found in Appendix \ref{ap:B}.

Having defined the dynamical structure factor, we can write the final form of the double differential cross-section as follows \cite{lovesey1984}
\begin{equation}
    \label{eq:ddcrossection}
    \frac{d^2\sigma}{d\Omega d\omega} = \frac{k_f}{k_i}(\gamma r_e)^2\sum_{jl}\left(\delta_{jl}-\frac{Q_jQ_l}{Q^2}\right)S_{jl}(\boldsymbol{Q},\omega)
\end{equation}
\noindent with $\gamma=+1.913$ being the dimensionless neutron gyromagnetic ratio and $r_e=e^2/m_ec^2$, the classical electron radius. An important consequence of Eq. \eqref{eq:ddcrossection} is that the poles in the susceptibility $\Pi_{im}(\boldsymbol{Q}, \omega)$ can be reflected as elementary excitations peaks in the scattering spectra.

\section{Magnetic Neutron Response in Prototypical Weyl Semimetals}\label{sec:prototypical}
The aim of this section is to compute the dynamical structure factor $S_{jl}(\boldsymbol{Q},\omega)$ for a toy model consisting of a single Weyl node using the formalism of Section \ref{sec:generic}. The low energy effective Hamiltonian for a single massless Weyl node is given by
\begin{equation}
    \label{eq:weylnodehamiltonian}
    H = \int d^3\boldsymbol{r}\ \sum_{\alpha\beta}\psi_{\alpha}^\dagger(\boldsymbol{r})\left[-i\hbar v_F \left(\boldsymbol{\sigma}^{\alpha\beta}\cdot\boldsymbol{\nabla}\right) -\mu\right]\psi_\beta(\boldsymbol{r})
\end{equation}
\noindent where $v_F$ is the isotropic Fermi velocity, $\boldsymbol{\sigma}=\left(\sigma_x,\sigma_y,\sigma_z\right)$ is the vector of Pauli matrices which operates on the pseudospin degrees of freedom, and $\mu$ is the bare chemical potential. Note that the eigenstates of $H$ have definite chirality. From Eq. \eqref{eq:weylnodehamiltonian}, the charge density operator is given by $\rho_{\boldsymbol{Q}}=-(e/V)\sum_{\boldsymbol{k}\alpha\beta}\psi_{\boldsymbol{k}\alpha}^\dagger\psi_{\boldsymbol{k}-\boldsymbol{Q}\beta}$ and the corresponding current operator is
\begin{equation}
    \label{eq:weylnodecurrent}
    \boldsymbol{J}_{\boldsymbol{Q}}=-\frac{ev_F}{V}\sum_{\boldsymbol{k}\alpha\beta}\psi_{\boldsymbol{k}\alpha}^\dagger\left[\boldsymbol{\sigma}^{\alpha\beta}
    \right]\psi_{\boldsymbol{k}-\boldsymbol{Q}\beta}
\end{equation}

Generally, the dynamical response function will depend on both the magnitude and the direction of the wavevector $\boldsymbol{Q}$. However, for isotropic Weyl nodes with rotational symmetry, we choose $\boldsymbol{Q}=Q\hat{\boldsymbol{z}}$ without loss of generality. Correspondingly, with this choice of $\boldsymbol{Q}$ direction, only the perpendicular components of the dynamical structure factor tensor $S_{xx}(\boldsymbol{Q},\omega)$ and $S_{yy}(\boldsymbol{Q},\omega)$ will contribute to the differential cross-section in Eq. \eqref{eq:ddcrossection}. Indeed, by using Eq. \eqref{eq:mdsf}, we obtain
\begin{equation}
    \label{eq:weylnodecrosssection}
    S_{jl}(\boldsymbol{Q},\omega)\propto\frac{1}{Q^2}\Big(\delta_{jl}\sum_i\Pi_{ii}(\boldsymbol{Q},\omega)-\Pi_{jl}(\boldsymbol{Q},\omega)\Big)
\end{equation}
from which we can see that only the components of the response function that are perpendicular to that of the dynamical structure factor need to be computed. For instance, only $\Pi_{xx}(\boldsymbol{Q},\omega)$ and $\Pi_{zz}(\boldsymbol{Q},\omega)$ components contribute to $S_{yy}(\boldsymbol{Q},\omega)$. Using Eq. \eqref{eq:magnetization} and the fact that $\boldsymbol{Q}$ is chosen is lie along the $\hat{\boldsymbol{z}}$ axis, $\Pi_{zz}(Q\hat{\boldsymbol{z}},\omega)$ does not contribute to the overall magnetization. Thus, for a component of the current along the $\hat{\boldsymbol{x}}$ axis (Fig. \ref{fig:1}), the magnetic neutron scattering signal will originate from $S_{yy}(Q\hat{\boldsymbol{z}}, \omega)$ which, for its part, only possesses a contribution from $\Pi_{xx}(Q\hat{\boldsymbol{z}}, \omega)$.

We explicitly compute the current-current correlation function $\Pi_{xx}(Q\hat{\boldsymbol{z}},\omega)$ from the effective low energy Hamiltonian in Eq. \eqref{eq:weylnodehamiltonian}. The corresponding non-interacting Green's function of a single Weyl node in the Matsubara frequency domain is given by
\begin{equation}
    \label{eq:greensfunctionweyl}
    G^{-1}(\boldsymbol{k},i\omega_n) = (i\omega_n+\mu)\mathcal{I}-\hbar v_F\boldsymbol{\sigma}\cdot\boldsymbol{k},
\end{equation}
\noindent where $\mathcal{I}$ is the $2\times2$ identity matrix. The non-interacting current-current correlation function is given as
\begin{equation}
\begin{split}
    \label{eq:pixxnoninteraction}
    \Pi_{xx}(Q&\hat{\boldsymbol{z}},i\omega_n)=-\frac{v_F^2e^2}{V\beta}\times\\
    &\sum_{\boldsymbol{k}\nu}\text{Tr}\Big[\sigma_xG(\boldsymbol{k},i\omega_\nu)\sigma_xG(\boldsymbol{k}+\boldsymbol{Q},i\omega_\nu+i\omega_n)\Big]
\end{split}
\end{equation}

By summing over the Matsubara frequency and performing analytical continuation, we arrive at 
\begin{equation}
\begin{split}
    \label{eq:postsummation}
    \Pi_{xx}&(Q\hat{\boldsymbol{z}},\omega)=-\frac{v_F^2e^2}{2V}\sum_{\boldsymbol{k}}\sum_{\substack{s=\pm\\s^\prime=\pm}}h_{ss^\prime}\times\\
    &\frac{n_F\big(s\hbar v_F|\boldsymbol{k}|-\mu\big)-n_F\big(s^\prime\hbar v_F|\boldsymbol{k}+\boldsymbol{Q}|-\mu\big)}{\hbar\omega+s\hbar v_F|\boldsymbol{k}|-s^\prime\hbar v_F|\boldsymbol{k}+\boldsymbol{Q}|+i\eta}
\end{split}
\end{equation}
\noindent where the summation is over the sign denoted by $s, s^\prime$; $\eta\rightarrow0$; $n_F(E)=(e^{\beta E}+1)^{-1}$ is the Fermi-Dirac distribution function, and the prefactor $h_{ss^\prime}$ is defined as
\begin{equation*}
    \label{eq:hss}
    h_{ss^\prime}=1+ss^\prime\ \frac{2k_x^2-\boldsymbol{k}\cdot(\boldsymbol{k} + \boldsymbol{Q})}{|\boldsymbol{k}||\boldsymbol{k}+\boldsymbol{Q}|}.
\end{equation*}

\begin{figure}[ht!]
	\centering
	\includegraphics[width=\linewidth]{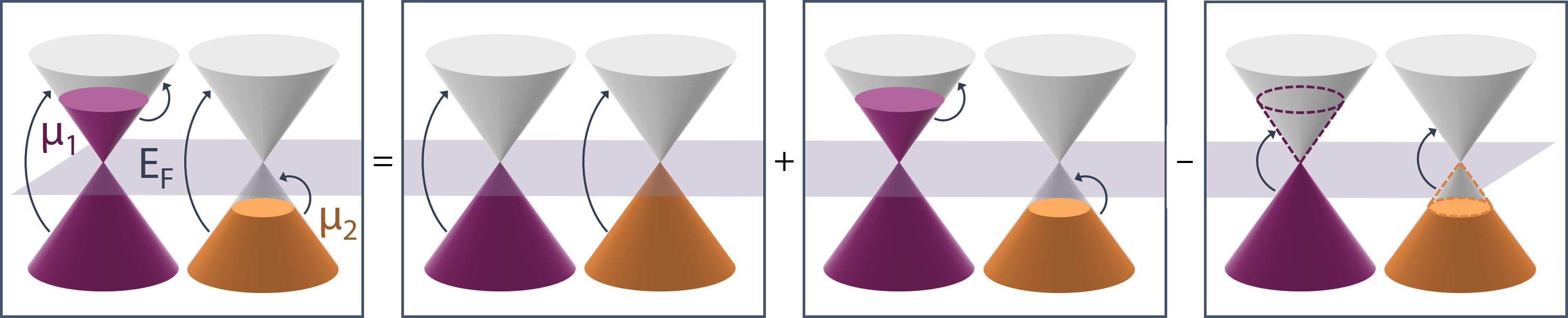}
	\caption{\textbf{Contributions to the response function.} Schematic of Weyl cones with chemical potentials $\mu_1$, $\mu_2$ above (purple) and below (orange) the Fermi level. The first contribution corresponds to interband transitions with $\mu=0$; the second, to intraband transitions in the upper (purple) and lower (orange) bands; and the third, as a negative contribution from forbidden transitions that arose from interband transitions when setting $\mu=0$.}
	\label{fig:2}
\end{figure}

To proceed, we first present the zero temperature calculation of Eq. \eqref{eq:postsummation} and leave the finite temperature result for Section \ref{sec:finiteT}. Only the imaginary component of the response function is presented as it contributes to the measured neutron signal. In this derivation, we focus on the case of $\omega>0$ without loss of generality as the $\omega<0$ case can be obtained by taking the complex conjugate (i.e., $\Pi(\boldsymbol{Q},-\omega) = \left[\Pi(\boldsymbol{Q},\omega)\right]^{*}$). The current-current correlation function can be decomposed into two separate pieces as
\begin{equation*}
    \Pi_{xx}(Q\hat{\boldsymbol{z}},\omega) = \Pi_{xx,\text{UD}}(Q\hat{\boldsymbol{z}},\omega) + \Pi_{xx,\text{D}}(Q\hat{\boldsymbol{z}},\omega)
\end{equation*}
corresponding to an undoped contribution $\Pi_{\text{UD}}(Q\hat{\boldsymbol{z}},\omega)$, where the chemical potential is located at the Weyl node, and a doped contribution $\Pi_{\text{D}}(Q\hat{\boldsymbol{z}},\omega)$, where the chemical potential is situated at a finite value away from the Weyl node. The current-current correlation function can be further separated into interband and intraband contributions as illustrated in Fig. \ref{fig:2} with the inclusion of a negative contribution stemming from extra, yet forbidden, transitions of the interband type \cite{wunsch2006,hwang2007}.

For the undoped portion, the contribution is solely from $s=-s^\prime=\pm1$ interband transitions and we obtain
\begin{equation}
    \label{eq:undoped}
    \text{Im}\big[\Pi_{xx,\text{UD}}\big]=\frac{e^2(\omega^2-v_F^2Q^2)}{24\pi\hbar v_F}\Theta\left(v_FQ+\omega\right)\Theta\left(-v_FQ+\omega\right)
\end{equation}
\noindent where $\Theta(x)$ denotes the Heaviside step function and the doped contribution is obtained as
\begin{equation}
    \label{eq:doped}
    \text{Im}\big[\Pi_{xx,\text{D}}\big]=-\frac{e^2(\omega^2-v_F^2Q^2)}{64\pi\hbar v_F}\Theta\left(Q^\prime+\omega^\prime\right)f\left(Q^\prime,\omega^\prime,\mu\right)
\end{equation}
\noindent where we introduce $Q^\prime=\hbar v_FQ$ and $\omega^\prime = \hbar\omega$ as arguments of a dimensionless function $f$ defined as
\begin{equation*}
\begin{split}
    \label{eq:f-function}
    &f(Q^\prime,\omega^\prime,\mu)=
    \sum_{s=\pm}\frac{8}{3}\Theta\left(-Q^\prime+\omega^\prime\right)\Theta\Big(s\mu-\frac{Q^\prime+\omega^\prime}{2}\Big)\\
    &+\sum_{s=\pm}s\Theta\left(-Q^\prime+\omega^\prime\right)\text{Rect}\Big(\frac{2\mu-s\omega^\prime}{Q^\prime}\Big)g^{(s)}\Big(\frac{2\mu-s\omega^\prime}{Q^\prime}\Big)\\
    &+\sum_{\substack{s=\pm\\s^\prime=\pm}}\Theta\left(Q^\prime-\omega^\prime\right)\Theta\Big(s\mu-\frac{Q^\prime+s^\prime\omega^\prime}{2}\Big)g^{(s^\prime)}\Big(\frac{-2ss^\prime\mu+\omega^\prime}{Q^\prime}\Big).
\end{split}
\end{equation*}
where the summation is over the sign denoted by $s, s^\prime$ and for conciseness, we define auxiliary functions $g^{(\pm)}$ and $\text{Rect}(x)$ as
\begin{equation*}
\begin{split}
    \label{eq:auxfunctions}
    g^{(\pm)}(x) &= x\Big(\frac{x^2}{3}+1\Big)\pm\frac{4}{3}\\
    \text{Rect}(x) = \Theta(1+x&)-\Theta(1-x)=\Theta(1-x)\Theta(1+x).
\end{split}
\end{equation*}
By substituting Eqs. \eqref{eq:undoped} and \eqref{eq:doped} into Eq. \eqref{eq:mdsf}, the dynamical structure factor can be written as
\begin{equation}
\begin{split}
    \label{eq:dsfforonenode}
    S&_{yy}(Q\hat{\boldsymbol{z}},\omega) = -\frac{e^2V}{24\pi^2(2\mu_Bc)^2}\frac{\omega^2-(v_FQ)^2}{(v_FQ)^2}v_F\times\\
    & \Big[\Theta\left(v_FQ+\omega\right)\Big(\Theta(-v_FQ+\omega)-\frac{3}{8}f(Q^\prime,\omega^\prime,\mu)\Big)\Big].
\end{split}
\end{equation}

\begin{figure}[ht!]
	\centering
	\includegraphics[width=0.99\linewidth]{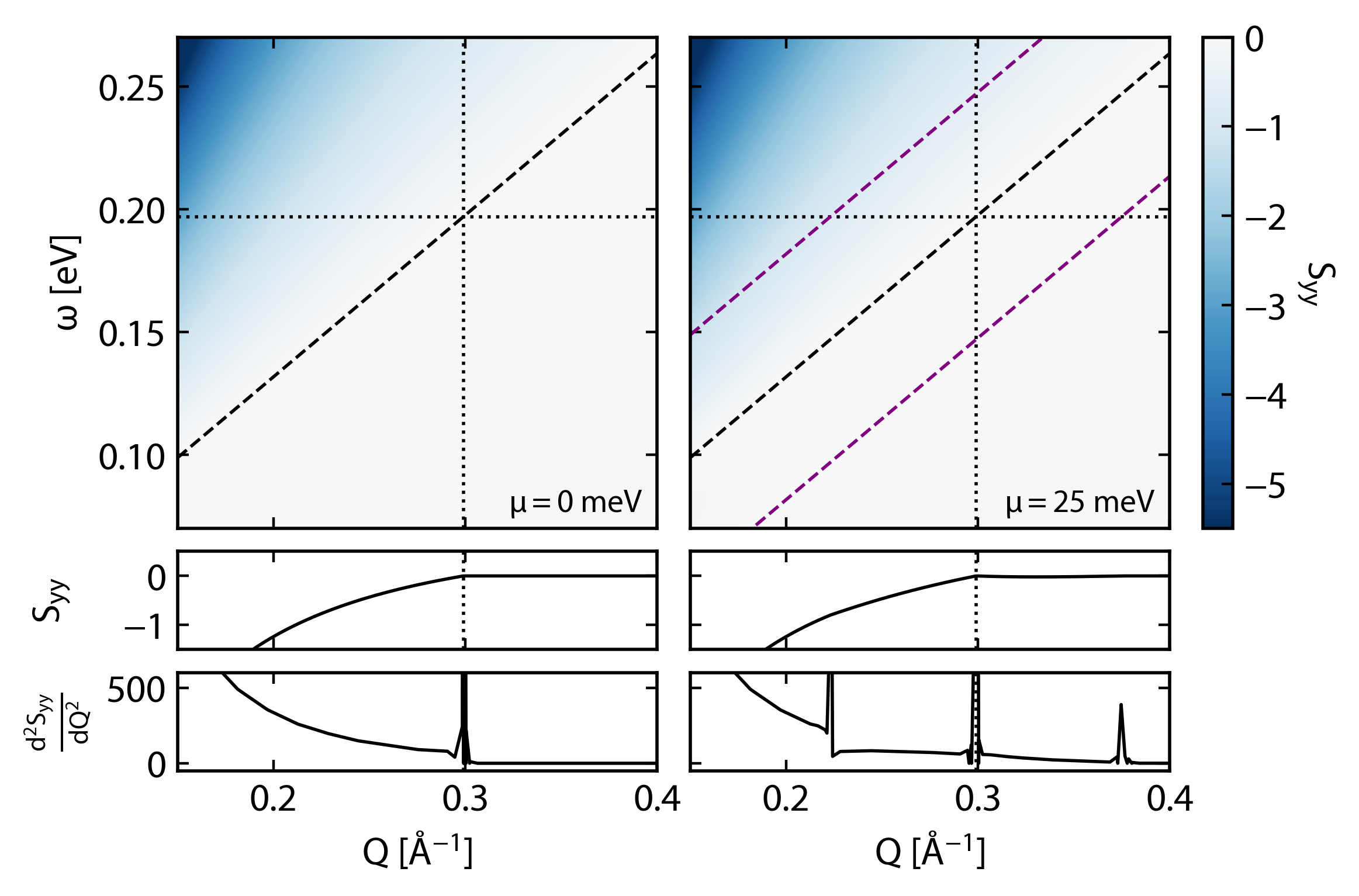}
	\caption{\textbf{Dynamical structure factor at zero temperature.} $S_{yy}$ (in arbitrary units) as a contour plot against wavevector $Q$ and frequency $\omega$ at zero (left) and finite (right) chemical potential taken to be 25 meV. Line cuts of $S_{yy}$ at a constant frequency are shown below the contour plots along with values of the second derivative with respect to $Q$. There is a large discontinuity in the second derivative located along $\omega = v_FQ$ (black dotted line) and additional ones along $\omega = v_FQ \pm 2\mu$ (purple dotted lines) at finite chemical potential, leading to an overall "triplet" signature.}
	\label{fig:3}
\end{figure}

This is valid for a single Weyl node; for multiple Weyl nodes, the dynamical structure factor is obtained by summing over the independent intra-node, interband contributions of each Weyl node. Eq. \eqref{eq:dsfforonenode} contains rich physical meaning as shown in Fig. \ref{fig:3}. First, the dependence of $S_{yy}$ on $v_F$ suggests that a larger Fermi velocity can contribute larger signal. Second, the term involving $\Theta(-v_FQ+\omega)$ indicates the presence of a large and discontinuous signal, here in the second derivative of $S_{yy}$ located along the line $\omega = v_FQ$. Such discontinuity signifies the physical onset for the interband electronic transition within the Weyl node. There are additional discontinuities along $\omega = v_FQ \pm 2\mu$ for finite chemical potential. This leads to a hallmark "triplet" signature equally spaced in both energy and wavevector space. It is also worthwhile mentioning that although the relation $\omega = v_FQ$ resembles the form of linear phonon dispersion, the neutron energy transfer $\omega$ and momentum transfer $Q$ are linked by the Fermi velocity $v_F$ of the Weyl fermions. This offers a clear signature as to how neutron scattering can probe the value of the Fermi velocity in a nodal system. Since a time-of-flight neutron scattering measurement spans a four-dimensional ($\boldsymbol{Q},\omega$) space, the neutron-based approach can readily determine the Fermi velocity vector for anisotropic Weyl cones. 

\section{Response function of a Nodal Line Semimetal}\label{sec:nodalline} The response function can be generalized for the case of a NLSM by adopting the Hamiltonian $H = \sum_{i} d_i\sigma_i$ ($i=1,2$), where $\sigma_i$ are the Pauli matrices and
\begin{equation}
\begin{split}
    \label{eq:d_components}
    d_1(\boldsymbol{k})&=\frac{1}{2m}\left(k_x^2+k_y^2-p_0^2\right)\\
    d_2(\boldsymbol{k})&=v_Fk_z.
\end{split}
\end{equation}
\noindent $m$ is the mass, $p_0$ is a constant, $k_i$ are momentum components and we define $d(\boldsymbol{k}) = \sqrt{\sum_i d_i^2(\boldsymbol{k})}$. We define the Green's function
\begin{equation}
    \label{eq:nodallineGreens}
    G^{-1}(\boldsymbol{k},i\omega_n) = (i\omega_n+\mu)\mathcal{I} - \sum_i d_i\sigma_i.
\end{equation}

We are interested in computing the correlation function of Eq. \eqref{eq:magneticsusceptibility}, using the expression for the Green's function for a NLSM (Eq. \eqref{eq:nodallineGreens}). After defining the resulting currents from the Hamiltonian in second quantization (Eq. \eqref{eqa:currents} in Appendix \ref{ap:C}), we compute the polarization operator, analogous to Eq. \eqref{eq:pixxnoninteraction}, as
\begin{equation}
\begin{split}
    \label{eq:responsefunctionnlsm}
    \Pi_{im}&\left(\boldsymbol{Q},i\omega_n\right) = -\frac{e^2}{V \beta} \sum_{\boldsymbol{k}\nu} \mathrm{Tr}\Big[v_i\Big(\boldsymbol{k}+ \frac{\boldsymbol{Q}}{2}\Big)\times\\
    & G\left(\boldsymbol{k},i\omega_\nu\right)v_m\Big(\boldsymbol{k}+ \frac{\boldsymbol{Q}}{2}\Big) G\left(\boldsymbol{k}+\boldsymbol{Q}, i\omega_n + i\omega_\nu\right)\Big]
\end{split}
\end{equation}
\noindent where $v_{x,y} = (1/m)(k_{x,y} + Q_{x,y}/2)$ and $v_z = v_F\sigma_y$. 

In Appendix \ref{ap:C}, we show a detailed calculation of the response function by substituting the explicit expression for the Green's function (Eq. \eqref{eq:nodallineGreens}) into the polarization operator (Eq. \eqref{eq:responsefunctionnlsm}) followed by performing the Matsubara summation and momentum integration. As shown in Eq. \eqref{eq:mdsf}, since we are interested in the neutron dynamical structure factor, we take the imaginary part of the response function $\text{Im}\left[\Pi_{xx}(\boldsymbol{Q},\omega)\right]$. This leads to four separate contributions corresponding to intraband and interband transitions within the valence and conduction bands from $\boldsymbol{k}$ to $\boldsymbol{k+Q}$ (Eq. \eqref{eqa:fourcontributionstoimaginary}).

Here, for simplicity, we compute only one component in $\text{Im}\left[\Pi_{xx}(\boldsymbol{Q},\omega)\right]$ for the case of zero Fermi level and zero temperature. Only one contribution out of the four, associated with the transition between states with energies $-d(\boldsymbol{k})$ and $d(\boldsymbol{k}+\boldsymbol{Q})$, survives (i.e., $n_F(-d(\boldsymbol{k}))=1$ and $n_F(d(\boldsymbol{k+Q}))=0$). We assume that $\boldsymbol{Q} = Q\hat{\boldsymbol{z}}$ such that only the $\Pi_{xx}$ component is under consideration, as was done previously. After the Matsubara summation, the resulting expression is
\begin{equation}
\begin{split}
    \label{eq:nodallineresponsezerofermi}
    \text{Im}&\left[\Pi_{xx}(Q\hat{\boldsymbol{z}},\omega)\right] = \pi e^2\int\frac{d^3\boldsymbol{k}}{(2\pi)^3}\delta\big(\omega - d(\boldsymbol{k}) - d(\boldsymbol{k+Q})\big)\\
    &\times\mathrm{Tr}\big[n_F(-d(\boldsymbol{k})) - n_F(d(\boldsymbol{k+Q}))\big]\\
    &\times v_i\left(\boldsymbol{k} + \frac{\boldsymbol{Q}}{2}\right)S^{(+)}(\boldsymbol{k})v_m\left(\boldsymbol{k} + \frac{\boldsymbol{Q}}{2}\right)S^{(-)}(\boldsymbol{k}+\boldsymbol{Q})
\end{split}
\end{equation}
where the $S^{(\pm)}(\boldsymbol{k})$ matrices are defined in Eq. \eqref{eqa:smatrices}.

From Eq. \eqref{eq:nodallineresponsezerofermi}, we can remove the $\delta$-function by taking an integral over $k_z$. By setting the argument of the $\delta$-function to zero, we can obtain expressions for $k_z$ and subsequently, for $d(\boldsymbol{k})$ and $d(\boldsymbol{k+Q})$ at this value of $k_z$. Setting the argument of the delta function to zero and solving it requires the condition that $\omega^2 - v_F^2Q^2 > 0$ thereby determining the limits of integration  as shown in Appendix \ref{ap:C}. We compute the remaining integral over $k_{\bot}$ with limits of integration that are defined by conditions imposed by the $\delta$-function in Eq. \eqref{eqa:fourcontributionstoimaginary} to obtain the response function for a NLSM at $\mu$, $T=0$ as
\begin{equation}
\begin{split}
    \label{eq:nodallineresponsezerofermizerotemp}
    \text{Im}&\left[\Pi_{xx}(Q\hat{\boldsymbol{z}},\omega)\right] =\ -\frac{e^2(\omega^2-v_F^2Q^2)}{16\pi \hbar v_F}\Theta\left(-v_FQ+\omega\right)\times\\
    &\Big[\frac{p_0^2\arcsin \sqrt{1-\xi}}{2m\sqrt{\omega^2-v_F^2Q^2}}-\frac{\xi}{2}\sqrt{1-\xi} +\frac{1}{3}\left(1-\xi\right)^{3/2}\Big]
\end{split}
\end{equation}
\noindent where we define an auxiliary function $\xi$ as
\begin{equation*}
\begin{split}
    \label{eq:xidefinition}
    \xi=\frac{p_0^4}{m^2\left(\omega^2-v_F^2Q^2\right)}.
\end{split}
\end{equation*}
The resulting neutron dynamical structure factor $S_{yy}(Q\hat{\boldsymbol{z}},\omega)$ given through the imaginary part of the response function of Eq. \eqref{eq:nodallineresponsezerofermizerotemp} also displays non-analytical behavior at $\omega = v_FQ$ resulting in a similar plot to Fig. \ref{fig:3} (with $\mu$ is taken to be zero). Analogous to Section \ref{sec:prototypical}, the effect of nonzero chemical potential in NLSMs will introduce additional discontinuities at $\omega = v_FQ \pm 2\mu$, which can revealed through a similar procedure of calculations. Ultimately, this captures the relationship between the energy transferred by the neutron $\omega$ and the Fermi velocity $v_F$ of the electronic quasiparticles within a NLSM. This link can be inferred through neutron scattering experiments to extract the value of $v_F$, much like the case of a simple WSM.

\section{Using Neutron Probes to Explore Weyl Nodes}\label{sec:locations} We proceed to calculate the dynamical structure factor for a realistic WSM with multiple pairs of Weyl nodes in the Brillouin zone. Without loss of generality, we assume the Weyl nodes to be located at momentum space locations $\boldsymbol{k}_{W,\zeta}$ with Fermi velocity $v_{F,\zeta}$ and chemical potential $\mu_{\zeta}$, where $\zeta=1,2$ labels the subset of Weyl node. The low-energy Hamiltonian for each Weyl node with explicit momentum space position follows from Eq. \eqref{eq:weylnodehamiltonian} with the shift $\boldsymbol{\nabla} \rightarrow \boldsymbol{\nabla}-i\boldsymbol{k}_{W,\zeta}$. Here we assume that the two Weyl nodes in consideration share the same chirality. The corresponding Matsubara Green's function for each Weyl node $G_\zeta(\boldsymbol{k},i\omega_n, \mu_\zeta)$ is
\begin{equation}
    \label{eq:matsubaraforeachnode}
    G_\zeta^{-1}(\boldsymbol{k}, i\omega_n, \mu_\zeta) = (i\omega_n+\mu_\zeta)\mathcal{I}-\hbar v_{F,\zeta}\boldsymbol{\sigma}\cdot\boldsymbol{k}_{\zeta}.
\end{equation}
This is the same result as Eq. \eqref{eq:greensfunctionweyl} with a shift due to the position of the Weyl node as $\boldsymbol{k}_\zeta=\boldsymbol{k}-\boldsymbol{k}_{W,\zeta}$ and with proper indexing. If intra-Weyl node scattering is solely considered, the calculated dynamical structure factor between two Weyl nodes is actually a sum of two contributions originating from taking into account one single Weyl node. Inter-Weyl node scattering may also occur between two Weyl nodes at distinct positions $\boldsymbol{k}_{W,1}$ and $\boldsymbol{k}_{W,2}$ (with chemical potentials $\mu_1$, $\mu_2$, respectively). In this circumstance, analogous to Eq. \eqref{eq:pixxnoninteraction}, the current-current response function for inter-Weyl node scattering is
\begin{equation}
\begin{split}
    \label{eq:internoderesponse}
    \Pi&_{xx}^{\text{int}}(Q\hat{\boldsymbol{z}},i\omega_n,\mu_1,\mu_2) =-\frac{v_F^2e^2}{V\beta}\times\\
    &\sum_{\boldsymbol{k}\nu}\text{Tr}\big[\sigma_xG_1\left(\boldsymbol{k}_1,i\omega_\nu,\mu_1\right)\sigma_xG_2\left(\boldsymbol{k}_2+\boldsymbol{Q}, i\omega_\nu+i\omega_n,\mu_2\right)\big]
\end{split}
\end{equation}
\noindent with $G_\zeta(\boldsymbol{k},i\omega_n, \mu_\zeta)$ defined in Eq. \eqref{eq:matsubaraforeachnode}. Here, as a proof of principle, we do not consider a specific interaction potential that can lead to the inter-node scattering, nor the effect from the finite relaxation time. This assumes that strong inter-Weyl node scattering can be determined from phenomena like the Kohn anomaly \cite{nguyen2020}. To clarify the experimental signature of inter-Weyl node scattering, we assume that the two Weyl nodes share the same isotropic Fermi velocity, i.e., $v_F = v_{F,1} = v_{F,2}$. Similar to the intra-Weyl node scattering of Section \ref{sec:prototypical}, the current-current response function of Eq. \eqref{eq:internoderesponse} can be split into an undoped ($\mu_1=\mu_2=0$) and a doped ($\mu_1, \mu_2\neq0$) contribution as
\begin{equation*}
    \label{eq:undoped-doped-weylnodes}
    \Pi_{xx}^{\text{int}}(\boldsymbol{Q},\omega,\mu_1,\mu_2)=\Pi_{xx,\text{UD}}^{\text{int}}(\boldsymbol{Q},\omega) + \Pi_{xx,\text{D}}^{\text{int}}(\boldsymbol{Q},\omega,\mu_1,\mu_2)
\end{equation*}
\begin{figure*}[ht!]
	\centering
	\includegraphics[width=\linewidth]{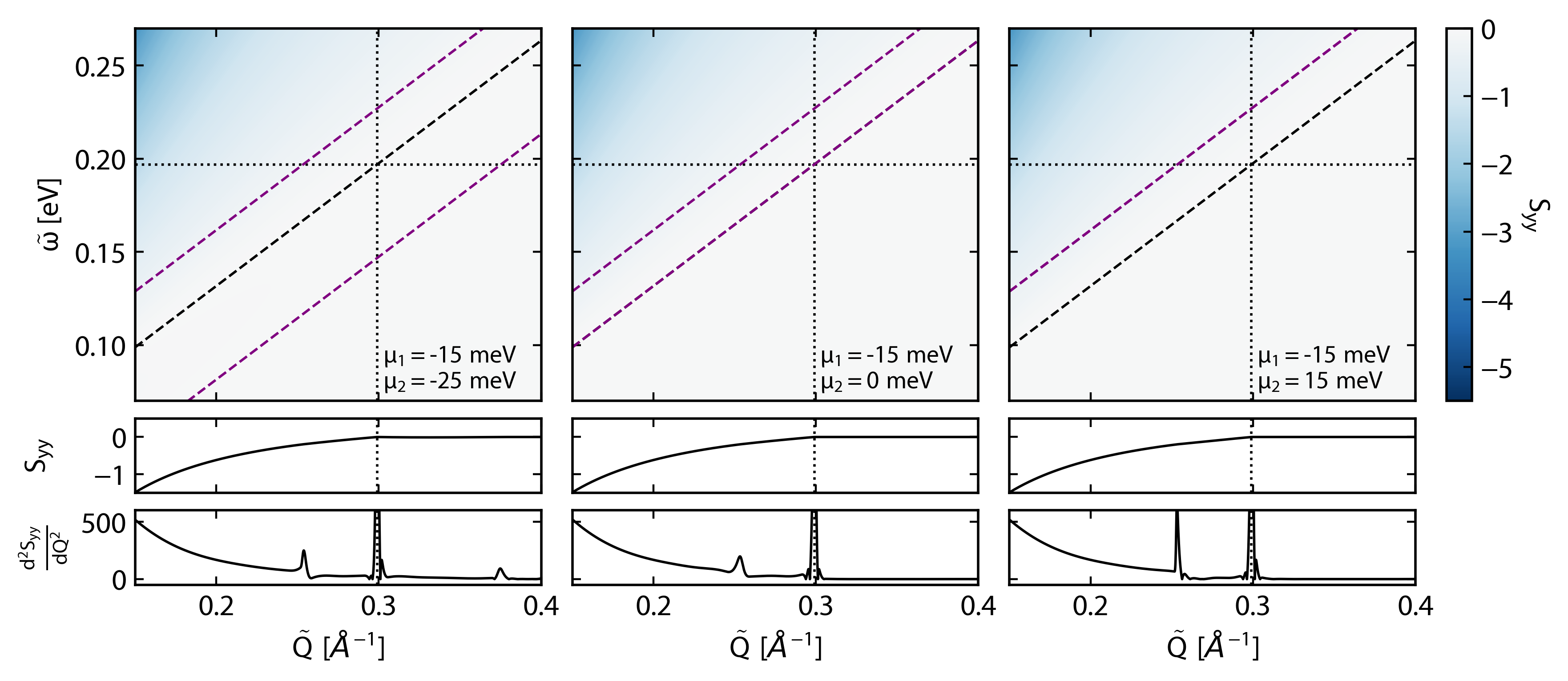}
	\caption{\textbf{Dynamical structure factor at zero temperature from inter-Weyl node scattering.} $S_{yy}$ (in arbitrary units) as a contour plot against $\tilde{Q}=\hbar v_F\left|\boldsymbol{Q} + \boldsymbol{k}_{W,1} - \boldsymbol{k}_{W,2}\right|$ and $\tilde{\omega}=\hbar\omega-\mu_1+\mu_2$. For the purposes of the plot, the chemical potential of the first Weyl node was fixed to be -15 meV whereas that of the second Weyl node was taken to be -25 meV (left), 0 meV (center) and 15 meV (right). Line cuts of $S_{yy}$ at constant $\tilde{\omega}$ are shown along with values of the second derivative with respect to $\tilde{Q}$. Dotted purple lines correspond to $\tilde{\omega} = \tilde{Q} - 2\mu_1$ and $\tilde{\omega} = \tilde{Q}+2\mu_2$ whereby, in addition to the line at $\tilde{\omega}=\tilde{Q}$ (dotted black line), the second derivative in $S_{yy}$ reveals discontinuous behavior.}
	\label{fig:4}
\end{figure*}
The imaginary part of the undoped contribution is computed as
\begin{equation*}
\begin{split}
    \label{eq:undoped-internode}
    \text{Im}&\left[\Pi_{xx,\text{UD}}^{\text{int}}(\boldsymbol{Q},\omega)\right]=\frac{e^2\big(\tilde{\omega}^2-\tilde{Q}^2\big)}{24\pi\hbar^3 v_F}\times\\
    &\Big[\Theta\big(\tilde{Q} + \tilde{\omega}\big)\Theta\big(-\tilde{Q}+\tilde{\omega}\big)-\Theta\big(\tilde{Q}-\tilde{\omega}\big)\Theta\big(-\tilde{Q}-\tilde{\omega}\big)\Big]
\end{split}
\end{equation*}
\noindent where $\tilde{Q}=\hbar v_F\left|\boldsymbol{Q}+\boldsymbol{k}_{W,1}-\boldsymbol{k}_{W,2}\right|$ represents the momentum transfer with the consideration of inter-Weyl node scattering and $\tilde{\omega}=\hbar\omega - \mu_1 + \mu_2$ is the associated energy transfer. Note that $\tilde{\omega}$ can be both positive and negative depending on the relative importance between the incoming neutron frequency and the chemical potentials of the Weyl nodes. If we define an auxiliary function $\tilde{f}$ (with associated auxiliary functions $g^{(\pm)}(x)$ and $\text{Rect}(x)$ defined previously) given by
\begin{equation*}
\begin{split}
    &\tilde{f}\big(\tilde{Q},\tilde{\omega},\mu_1,\mu_2\big) = \frac{8}{3}\Theta\big(\tilde{Q}-\tilde{\omega}\big)\Theta\big(-\tilde{Q}-\tilde{\omega}\big)\Theta\Big(\frac{\tilde{\omega}+2\mu_1}{\tilde{Q}}-1\Big)\\
    &- \frac{8}{3}\Theta\big(\tilde{Q}+\tilde{\omega}\big)\Theta\big(-\tilde{Q}+\tilde{\omega}\big)\Theta\Big(\frac{-\tilde{\omega}+2\mu_2}{\tilde{Q}}-1\Big)\\
    &+ g^{(+)}\Big(\frac{\tilde{\omega}+2\mu_1}{\tilde{Q}}\Big)\Theta\big(\tilde{Q}-\tilde{\omega}\big)\Theta\big(-\tilde{Q}-\tilde{\omega}\big)\text{Rect}\Big(\frac{\tilde{\omega}+2\mu_1}{\tilde{Q}}\Big)\\
    &- g^{(-)}\Big(\frac{\tilde{\omega}+2\mu_1}{\tilde{Q}}\Big)\Theta\big(\tilde{Q}+\tilde{\omega}\big)\Theta\big(\tilde{Q}-\tilde{\omega}\big)\Theta\Big(\frac{\tilde{\omega}+2\mu_1}{\tilde{Q}}-1\Big)\\
    &+ g^{(-)}\Big(\frac{\tilde{\omega}-2\mu_2}{\tilde{Q}}\Big)\Theta\big(\tilde{Q}+\tilde{\omega}\big)\Theta\big(-\tilde{Q}+\tilde{\omega}\big)\text{Rect}\Big(\frac{\tilde{\omega}-2\mu_2}{\tilde{Q}}\Big)\\
    &- g^{(+)}\Big(\frac{\tilde{\omega}-2\mu_2}{\tilde{Q}}\Big)\Theta\big(\tilde{Q}-\tilde{\omega}\big)\Theta\big(\tilde{Q}+\tilde{\omega}\big)\Theta\Big(\frac{-\tilde{\omega}+2\mu_2}{\tilde{Q}}-1\Big)
\end{split}
\end{equation*}
then the imaginary part of the doped contribution is
\begin{equation*}
\begin{split}
    \label{eq:doped-internode}
    \text{Im}&\left[\Pi_{xx,\text{D}}^{\text{int}}(\boldsymbol{Q},\omega,\mu_1,\mu_2)\right]=\frac{e^2\big(\tilde{\omega}^2-\tilde{Q}^2\big)}{64\pi\hbar^3 v_F}\times\\
    &\Big(\tilde{f}\big(\tilde{Q},\tilde{\omega},\mu_1,\mu_2\big)-\tilde{f}\big(\tilde{Q},-\tilde{\omega},-\mu_1,-\mu_2\big)\Big).
\end{split}
\end{equation*}

The first $\tilde{f}$ function in the expression for $\text{Im}\left[\Pi_{xx, D}^{\text{int}}(\boldsymbol{Q},\omega,\mu_1,\mu_2)\right]$ produces a nonzero contribution only when the condition $\mu_1,\mu_2>0$ is satisfied, whereas the second $\tilde{f}$ function is nonzero only for $\mu_1,\mu_2 <0$. These two terms correspond to the electron and the hole responses, respectively.

The corresponding structure factor $S_{yy}(Q\hat{\boldsymbol{z}},\omega)$ originating from both undoped and doped contributions of the neutron response function for the case of inter-Weyl node scattering, at zero temperature, is computed as
\begin{equation}
\begin{split}
    \label{eq:internodestructurefactor}
    S&_{yy}(Q\hat{\boldsymbol{z}},\omega) = -\frac{e^2V}{24\pi^2(2\mu_Bc)^2}\frac{\tilde{\omega}^2-\tilde{Q}^2}{\tilde{Q}^2}v_F\times\\
    & \Big[\Theta\big(\tilde{Q}+\tilde{\omega}\big)\Theta\big(-\tilde{Q}+\tilde{\omega}\big)-\Theta\big(\tilde{Q}-\tilde{\omega}\big)\Theta\big(-\tilde{Q}-\tilde{\omega}\big)\\
    & + \frac{3}{8}\big(\tilde{f}(\tilde{Q},\tilde{\omega},\mu_1,\mu_2) - \tilde{f}(\tilde{Q},-\tilde{\omega},-\mu_1,-\mu_2)\big)\Big].
\end{split}
\end{equation}

Due to the presence of terms involving the Heaviside function in Eq. \eqref{eq:internodestructurefactor}, including those in $\tilde{f}$, one expects a large non-analytical neutron signal at $\tilde{Q}=\tilde{\omega}$ or, equivalently, at $\hbar v_F\boldsymbol{Q} = \pm\hbar v_F\left(\boldsymbol{k}_{W,2}-\boldsymbol{k}_{W,1}\right) + \tilde{\omega}$ as a result of inter-Weyl node scattering, as seen in Fig. \ref{fig:4}. Since one usually maintains control over $\tilde{\omega}$ through a priori knowledge of the chemical potential of the Weyl nodes and selection of the incoming neutron energy, the measured neutron signal can serve as a probe for the momentum space location of the distinct Weyl nodes within the material system. Measuring the neutron signal with suitable instruments such as triple-axis and time-of-flight neutron spectroscopies followed by performing derivatives of the signal with respect to $\boldsymbol{Q}$ reveals the non-analytical behavior. Since the latter forms along the line $\boldsymbol{Q} \approx \boldsymbol{k}_{W,2}-\boldsymbol{k}_{W,1}$, it is readily discernible in the neutron data and provides information about the different nesting conditions between the nodes that stem from the material system, thereby revealing their momentum space locations.

\section{Signature of Chiral Anomaly in Neutron Spectra}\label{sec:chiralanomaly}
In this section, we focus on an important phenomenon associated with WSMs which is the chiral anomaly \cite{nielsen1983,zyuzin2012,son2013,chang2015}. It is observed when a pair of parallel electric and magnetic fields, $\boldsymbol{E}$ and $\boldsymbol{B}$, induces a non-local pumping from one node possessing positive chirality to another with negative chirality for $\boldsymbol{E}\cdot\boldsymbol{B}>0$ (and vice-versa for $\boldsymbol{E}\cdot\boldsymbol{B}<0$). The charge transfer is stabilized by an inter-Weyl node scattering mechanism with characteristic relaxation time given by $\tau$. The low-energy Hamiltonian in the vicinity of a Weyl node is given by Eq. \eqref{eq:weylnodehamiltonian}, but unlike Section \ref{sec:locations}, we reintroduce chirality $\lambda$ along with the shift in momentum position from the origin of the Weyl nodes by taking $\boldsymbol{\nabla} \rightarrow \boldsymbol{\nabla}-i\lambda\boldsymbol{k}_{W}$ and $\mu\rightarrow\mu_\lambda$ with $\lambda = \pm 1$. The chiral chemical potential $\mu_\lambda$ is given by \cite{zhou2015, thakur2018}
\begin{equation}
    \label{eq:chiralmu}
    \mu_\lambda = \big(\mu^3 + \lambda\frac{3e^2\hbar v_F^3}{2c}\tau\boldsymbol{E}\cdot\boldsymbol{B}\big)^{\frac{1}{3}}
\end{equation}
\noindent where $\mu$ is the chemical potential without any consideration of chirality and $\tau$ is the internode scattering timescale. Eq. \eqref{eq:chiralmu} is only valid in the limit of weak magnetic field whereby the discreteness of Landau levels can be ignored. The current-current response function in WSMs in the context of the chiral anomaly has been previously studied in Ref. \cite{thakur2018}. It is mentioned that for the chiral anomaly to be observable in experiments, the shift in the chemical potential seen in Eq. \eqref{eq:chiralmu} should be on the order of the chemical potential $\mu$ itself. By taking typical values of $v_F$ and $\mu$ for WSMs, it is shown that the required value of $\boldsymbol{E}\cdot\boldsymbol{B}$ for observation is attainable in present-day experiments. Here, we follow a similar approach, but apply the results towards the calculation of the dynamical structure factor for neutron scattering experiments. The corresponding Matsubara Green's function in the vicinity of the Weyl node with chirality $\lambda$ is given by Eq. \eqref{eq:greensfunctionweyl} with $\boldsymbol{k}\rightarrow\boldsymbol{k}_{\lambda}=\boldsymbol{k}+\lambda\boldsymbol{k}_W$. The response function for two Weyl nodes with opposite chirality can be obtained by simply summing the contributions from the two Weyl nodes with a modified chemical potential $\mu\rightarrow\mu_\lambda$. The imaginary part of the current-current response function can be found in terms of this modified chemical potential. The corresponding dynamical structure factor at zero temperature is obtained as
\begin{equation}
\begin{split}
    \label{eq:dsfforchiralanomaly}
    S&_{yy}(Q\hat{\boldsymbol{z}},\omega) = -\frac{e^2V}{24\pi^2(2\mu_Bc)^2}\frac{\omega^2-(v_FQ)^2}{(v_FQ)^2}v_F\times\\
    &\Big[\sum_{\lambda=\pm1}\Theta\left(v_FQ+\omega\right)\big(\Theta(-v_FQ+\omega)-\frac{3}{8}f(Q',\omega',\mu_\lambda)\big)\Big].
\end{split}
\end{equation}

\begin{figure}[ht!]
	\centering
	\includegraphics[width=\linewidth]{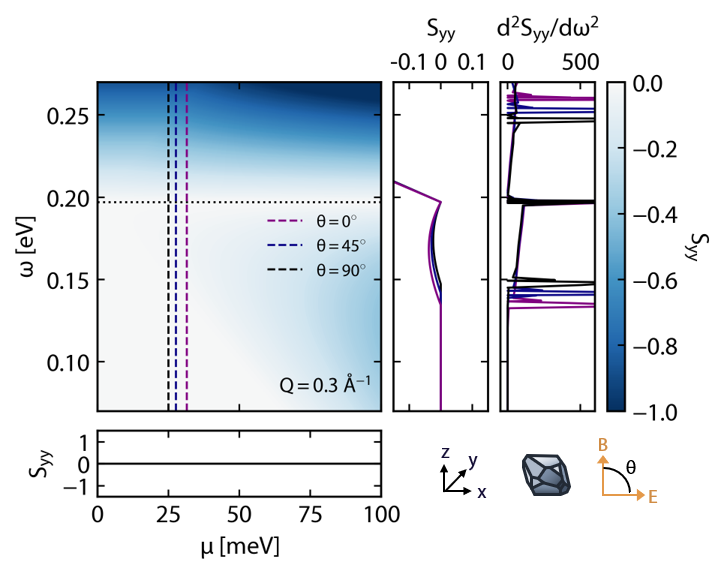}
	\caption{\textbf{Signatures of chiral anomaly in WSMs through the dynamical structure factor.} $S_{yy}$ (in arbitrary units) as a contour plot against chemical potential $\mu_\lambda$ of the Weyl node and frequency $\omega$ at a fixed wavevector $Q = 0.3$ \si{\angstrom}$^{-1}$. Associated line cuts at fixed values of $\omega$ (bottom) and $\mu_\lambda$ (right) are shown. Notably, three different line cuts at fixed $\mu_\lambda$ correspond to different angles between applied electric $\boldsymbol{E}$ and magnetic $\boldsymbol{B}$ fields defined by $\theta$ = 0$^\circ$ (black), 45$^\circ$ (blue), and 90$^\circ$ (purple). These values of $\mu_\lambda$ correspond to a shift from the original value $\mu$ induced by the chiral anomaly as implied by the $\boldsymbol{E}\cdot\boldsymbol{B}$ term in Eq. \eqref{eq:chiralmu}. The $\boldsymbol{E}\cdot\boldsymbol{B}$-induced shift is taken to be on the same order as $\mu^3$ where $\mu$ is set to 25 meV and positive chirality is assumed. The right-most plot shows the second derivative of $S_{yy}$ with respect to the frequency $\omega$ to highlight divergent behavior at $\omega = v_FQ$ and $\omega = v_FQ \pm 2\mu_\lambda$, the latter of which is heavily influenced by the $\boldsymbol{E}\cdot\boldsymbol{B}$ shift.}
	\label{fig:5}
\end{figure}

Similar to the discussion in Section \ref{sec:prototypical}, at a constant value of wavevector (which is a common experimental measurement for triple-axis neutron spectroscopy), the second derivative in the dynamical structure factor $S_{yy}$ with respect to the frequency $\omega$ displays divergence-like behavior at $\omega=v_FQ$ and at $\omega=v_FQ \pm 2\mu_\lambda$. In this case, the chiral anomaly shifts the chemical potential from $\mu\rightarrow\mu_\lambda$ in accordance with Eq. \eqref{eq:chiralmu} due to an extra term that depends on $\boldsymbol{E}\cdot\boldsymbol{B}$. Hence upon application of parallel electric and magnetic fields, one expects that the divergence in the second derivative will also be shifted and occur at $\omega=v_FQ \pm 2\mu_\lambda$ whereas the divergence at $\omega=v_FQ$ remains unaltered. This behavior is illustrated in Fig. \ref{fig:5} and serves, by itself, as a novel experimental signature of the chiral anomaly in WSMs. Provided that the shift $\mu\rightarrow\mu_\lambda$ is on the order of the original value of the chemical potential, the shift in $\omega$ whereupon the divergence in the second derivative occurs can be measured as a function of the angle $\theta$ between $\boldsymbol{E}$ and $\boldsymbol{B}$ to unveil information on $v_F$ and remarkably, on $\tau$, the latter of which is generally difficult to quantify in experiment.

\section{Massive Dirac Fermions}\label{sec:finitemass} 
We investigate how introducing a finite mass to the massless Weyl fermions of Eq. \eqref{eq:weylnodehamiltonian} induces a change to the dynamical structure factor. It is convenient to rewrite Eq. \eqref{eq:weylnodehamiltonian} into a massive four-band Hamiltonian in terms of the anti-commuting four-component $\gamma$ matrix \cite{kaku1993}
\begin{equation}
    \label{eq:massive-four-band-Hamiltonian}
    H=\int d^3\boldsymbol{r}\ \bar{\psi}(\boldsymbol{r})\big[ -iv_F\left(\boldsymbol{\gamma}\cdot\boldsymbol{\nabla}\right)
    +m+\slashed{k_w}\gamma^5-\mu\gamma^0\big]\psi(\boldsymbol{r})
\end{equation}
\noindent where $\bar{\psi}\equiv\psi^\dagger\gamma^0$ is the Dirac conjugate spinor field, $\boldsymbol{\gamma}=(\gamma^1,\gamma^2,\gamma^3)$, $\gamma^5 = i\gamma^0\gamma^1\gamma^2\gamma^3$, and $(k_w)^\mu=(k_w^0,\boldsymbol{k}_w)$ is a 4-vector. In this derivation, we set $\hbar=v_F=1$ for simplicity and employ the Feynman slash notation, wherein $\slashed{W}=\gamma_\mu W^\mu$ for the $W^\mu$ 4-vector. We see that $\bar{\psi}\slashed{k}_w\gamma^5\psi$ term leads to an induced Chern-Simons term of the form $(k_w)_\mu\epsilon^{\mu\nu\lambda\sigma}F_{\lambda\sigma}A_{\nu}$ in a WSM \cite{assuncao2015,zyuzin2012}. The chiral shift $\boldsymbol{k}_w$ is present in the free Hamiltonian of WSMs. Additionally, this shift can also be dynamically generated in DSMs in the normal phase with the presence of a magnetic field \cite{kim2013}. Similar to the case of graphene, the generation of the Chern-Simons term implies a topological nature of the normal state in this material \cite{redlich1985}. We compute the current operators of the Hamiltonian in Eq. \eqref{eq:massive-four-band-Hamiltonian} to obtain
\begin{equation}
\begin{split}
    \label{eq:currentoperator-fourband}
    J_\mu &= e\bar{\psi}\gamma_\mu\psi\\
    J_\mu^5 &= e\bar{\psi}\gamma_\mu\gamma^5\psi.
\end{split}
\end{equation}

The first current operator in Eq. \eqref{eq:currentoperator-fourband} is known as the gauge current with global U(1) gauge invariance and the second, an axial (chiral) current operator. As a result of the chiral magnetic effect, the axial current density $J_\mu^5$ is generated in the free theory when a fermion charge density and a magnetic field are present \cite{newman2006}. From the expression of $J_\mu^5$, the chiral shift $\boldsymbol{k}_w$ is already induced to the lowest order in perturbation theory even if $k_w^0 = 0$. As a result, a DSM with a nonzero charge density is transformed into a WSM immediately when an external magnetic field is applied to the system.

We proceed to calculate the current-current response function using the relativistic-like model of Eq. \eqref{eq:massive-four-band-Hamiltonian} and the current operators of Eq. \eqref{eq:currentoperator-fourband}. This response function is written in relativistic notation \cite{kaku1993} as
\begin{equation}
    i\Pi_{im}(Q) = -e^2\int\frac{d^4k}{(2\pi)^4}\ \text{Tr}\left[\gamma_iG(k)\gamma_mG(k+Q)\right]
\end{equation}
\noindent where $iG^{-1}(k)=\slashed{k}-\slashed{k_w}\gamma^5-m$ and $\int d^4k/(2\pi)^4$ denotes an integral over the four-dimensional Euclidean momentum space.

When we rewrite the propagator $G^{-1}(k)$ in terms of $k_w^\mu$, we obtain three contributions: the first one is proportional to the vacuum polarization of quantum electrodynamics, while the second and third ones are the contributions that are linear and quadratic in $k_w^\mu$, respectively. Although nonlinear terms may introduce further interesting phenomena, we restrict our analysis to the term that is linear in $k_w^\mu$ such that
\begin{equation}
    \label{eq:perturbPi}
    i\Pi_{im}(Q) = i\Pi_{im}^0(Q) + i\Pi_{im}^1(Q).
\end{equation}

The $\Pi_{im}^0(Q)$ is the usual vacuum polarization tensor of quantum electrodynamics \cite{kapusta2006}. The $\Pi_{im}^1$ term that is linear in $k_w^\mu$ is given by
\begin{equation}
\begin{split}
    i\Pi_{im}^1(Q)=-\int\frac{d^4k}{(2\pi)^4}\ \text{Tr}&\left[\gamma_iG_0(k)\gamma_mG_1(k+Q)\right.\\
    &\left.+\gamma_iG_1(k)\gamma_mG_0(k+Q)\right]
\end{split}
\end{equation}
\noindent where $iG_0^{-1}(k)=\slashed{k}-m$ is the bare propagator at $k_w^\mu=0$ and $G_1(k)=\frac{i}{\slashed{k}-\slashed{k_w}\gamma^5-m}\slashed{k_w}\gamma^5G_0(k)$. We compute the imaginary part of the $\Pi_{im}^0$ term in the whole domain of $Q^2$ following analytical continuation \cite{kapusta2006} as
\begin{equation}
    \label{eq:imaginarypart-pizero}
    \text{Im}\left[\Pi_{im}^0(Q)\right]=-\left(Q_iQ_m-\eta_{im}Q^2\right)R(Q)\Theta\left(Q^2-4m^2\right)
\end{equation}
\noindent where $R(Q)$ is defined as
\begin{equation*}
    R(Q) = \frac{e^2}{12\pi}\frac{-8m^4-2m^2Q^2+Q^4}{Q^3\sqrt{Q^2-4m^2}},
\end{equation*}
 $Q^2 = \omega^2-|\boldsymbol{Q}|^2$ is the squared norm, and $\eta_{im}$ is the metric tensor. In Eq. \eqref{eq:imaginarypart-pizero}, we considered the case corresponding to $\mu=0$ and the dimensional regularization scheme was implemented \cite{thooft1972}. For $Q^2 < 4m^2$, the imaginary part of the vacuum polarization function is zero. The current-current response function for the undoped case in Section \ref{sec:prototypical}, shown in Eq. \eqref{eq:undoped}, corresponds to the polarization bubble diagram of quantum electrodynamics in Eq. \eqref{eq:imaginarypart-pizero} with $m=0$. We calculate the imaginary part of $\Pi_{im}^1$ to first order in $k_w^\mu$ as \cite{jackiw1999}
 \begin{equation}
    \label{eq:imaginarypart-pione}
     \text{Im}\left[\Pi_{im}^1(Q)\right] = -\epsilon_{imab}\frac{e^2m^2}{\pi}\frac{(k_w)^aQ^b}{Q\sqrt{Q^2-4m^2}}.
 \end{equation}
 
 We can write the dynamical structure factor at zero temperature using Eqs. \eqref{eq:mdsf}, \eqref{eq:imaginarypart-pizero} and \eqref{eq:imaginarypart-pione} to obtain
 \begin{equation}
 \begin{split}
     S_{jl}(\boldsymbol{Q},\omega)=-&\frac{1}{\pi}\left(\frac{1}{2\mu_B c}\right)^2\sum_{ikmn}\epsilon_{jik}\epsilon_{lmn}\frac{Q_kQ_n}{Q^4}\times\\
     &\left(\text{Im}\left[\Pi_{im}^0(Q)\right]+\text{Im}\left[\Pi_{im}^1(Q)\right]\right).
\end{split}
\end{equation}

\section{Effect of Finite Temperature}\label{sec:finiteT}
\begin{figure}[ht!]
	\centering
	\includegraphics[width=\linewidth]{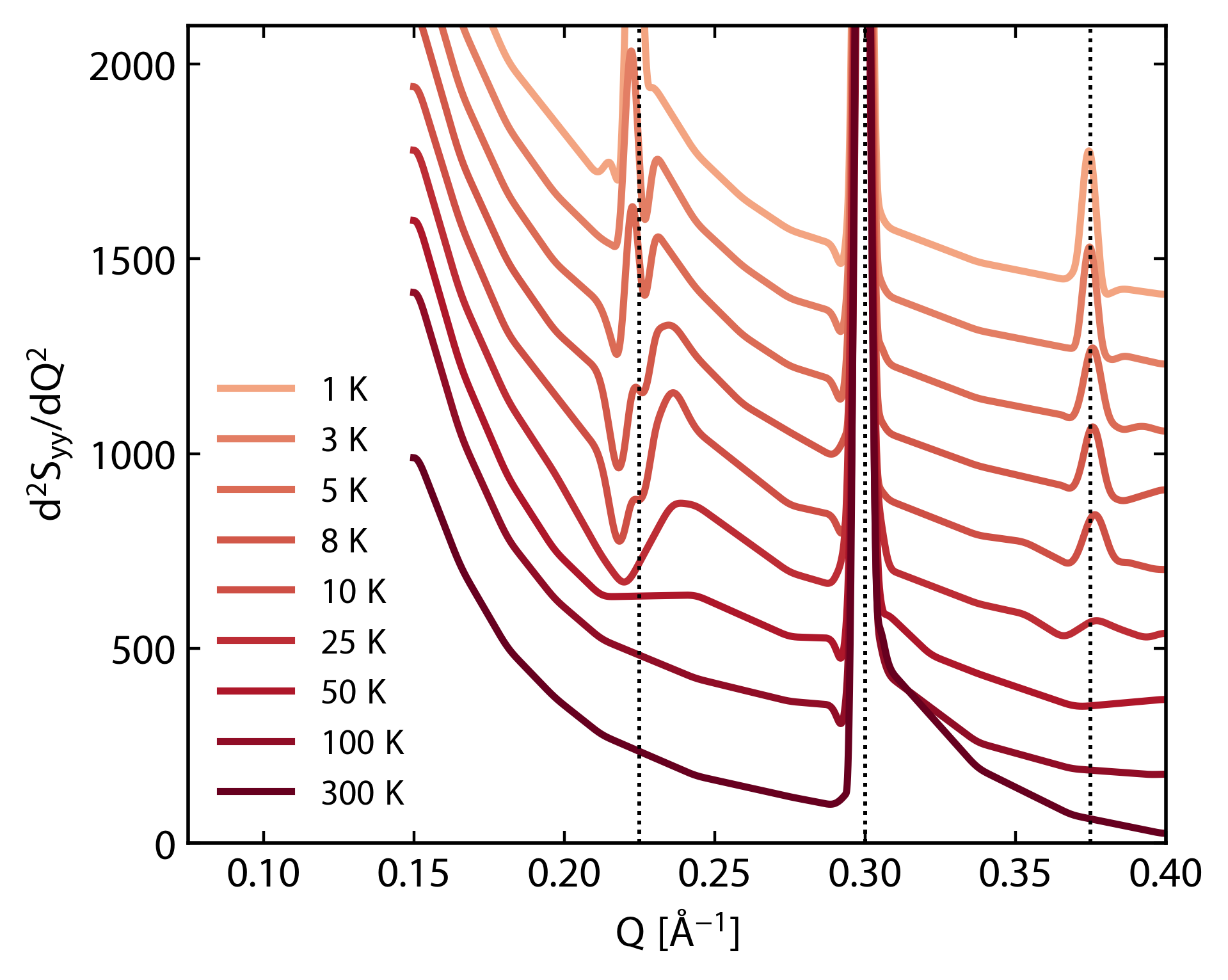}
	\caption{\textbf{Temperature dependence of second derivative.} Second derivative of $S_{yy}$ (in arbitrary units) with respect with wavevector $Q$ at different temperatures. The chemical potential is taken to be 25 meV. Each line plot corresponding to a temperature is vertically offset by 175 for clarity. The divergence in the second derivative at $\omega = v_FQ \pm 2\mu$ (left and right dotted lines) decreases with increase of temperature, whereas the signal at $\omega = v_FQ$ (central dotted line) prevails up to room temperature.}
	\label{fig:7}
\end{figure}

We calculate the dynamical structure factor of a Weyl node through the current-current response function with consideration of finite temperature. For intra-Weyl node scattering (Section \ref{sec:prototypical}), the charge neutral (undoped) part of the response function $\Pi_{xx,\text{UD}}(\boldsymbol{Q},\omega)$ remains unaffected whereas the doped contribution $\Pi_{xx,\text{D}}(\boldsymbol{Q},\omega)$ is modified by finite temperature. From the Matsubara summation of Eq. \eqref{eq:greensfunctionweyl}, we obtain the finite temperature expression
\begin{equation}
\begin{split}
    \label{eq:finite-temperature-response}
    \text{Im}&\left[\Pi_{xx,\text{D}}(\boldsymbol{Q},\omega)\right]=-\frac{e^2(\omega^2-v_F^2Q^2)}{64\pi\hbar v_F}\Theta(\omega+v_FQ)\\
    &\times\big[\Theta(\omega-v_FQ)H+\Theta(-\omega+v_FQ)\sum_{s=\pm}sG_s\big]
\end{split}
\end{equation}
\noindent with auxiliary functions $G_{\pm}$ and $H$ defined as
\begin{equation*}
\begin{split}
    G_{\pm}&=\sum_{s^\prime=\pm}\int_1^\infty du\ \frac{u^2+1}{\text{exp}\big(\frac{\hbar v_FQu \mp \hbar\omega+2s^{\prime}\mu}{2k_BT}\big)+1}\\
    H&=\sum_{s^\prime=\pm}\int_{-1}^1 du\ \frac{u^2+1}{\text{exp}\big(\frac{\hbar v_FQu + \hbar\omega+2s^\prime\mu}{2k_BT}\big)+1}.
\end{split}
\end{equation*}
\begin{figure*}[ht!]
	\centering
	\includegraphics[width=\linewidth]{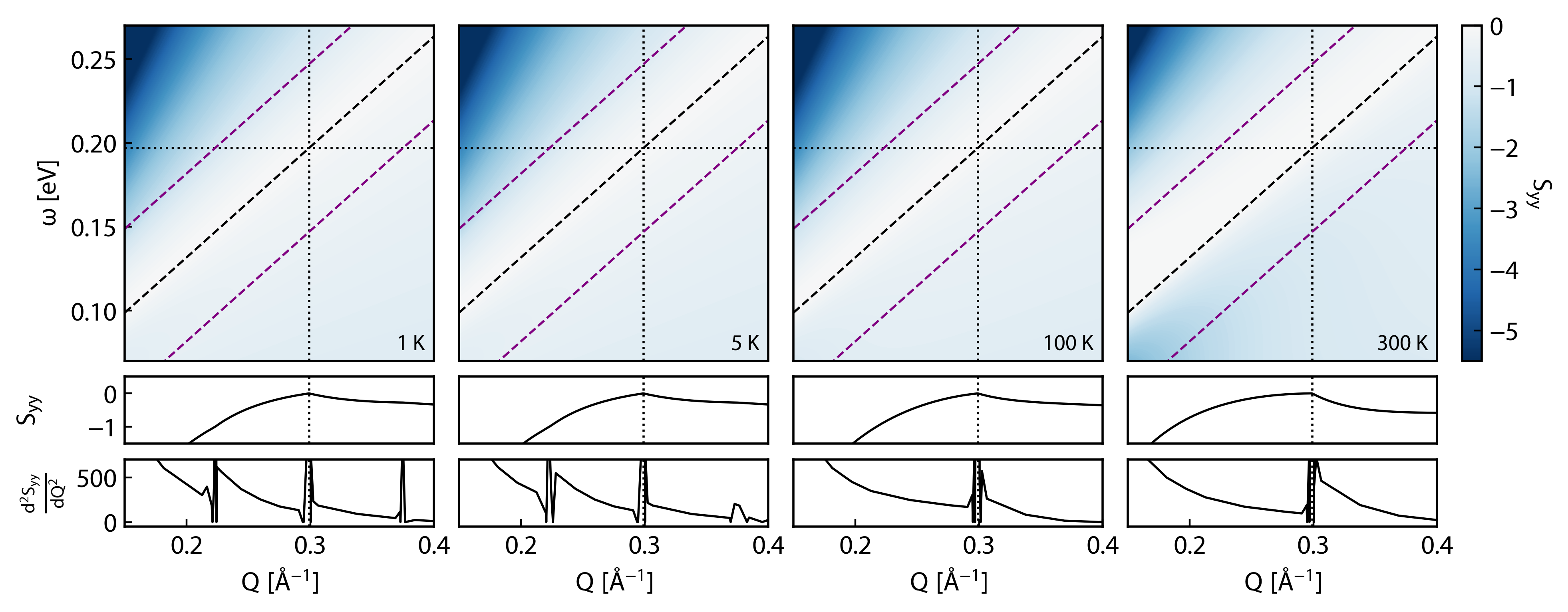}
	\caption{\textbf{Dynamical structure factor at finite temperature.} $S_{yy}$ (in arbitrary units) as a contour plot against wavevector $Q$ and frequency $\omega$ at different temperatures of 1K, 5K, 100K, and 300K. The chemical potential is taken to be 25 meV. Line cuts of $S_{yy}$ at a constant frequency are shown below these contour plots along with values of the second derivative with respect to $Q$. There is a large and discontinuity in the second derivative located along $\omega = v_FQ$ (black dotted line). The discontinuities associated with chemical potential at $\omega = v_FQ \pm 2\mu$ (purple dotted lines) are smeared out by the effect of finite temperature with a small remnant signal upwards to 10K (as suggested by the two left-most subplots).}
	\label{fig:6}
\end{figure*}
In particular, it can be shown that under the $T\rightarrow0$ limit, Eq. \eqref{eq:finite-temperature-response} reduces to the zero temperature result of Eq. \eqref{eq:doped}. The final form of the dynamical structure factor at finite temperature can be written as
\begin{equation}
\begin{split}
    \label{eq:finite-structure-factor}
    S&_{yy}(Q\hat{\boldsymbol{z}},\omega,T)=-\frac{e^2V}{24\pi^2(2\mu_Bc)^2}\frac{1}{1-e^{-\beta\hbar\omega}}\times\\
    & \frac{\omega^2-(v_FQ)^2}{(v_FQ)^2}v_F\Theta\left(v_FQ + \omega\right)\Big[\Theta(-v_FQ+\omega)-\\
    & \frac{3}{8}\big(\Theta(v_FQ-\omega)\sum_{s=\pm}sG_s+\Theta(-v_FQ+\omega)H\big)\Big].
\end{split}
\end{equation}
The effect of finite temperature from Eq. \eqref{eq:finite-structure-factor} is shown in Fig. \ref{fig:6}. As mentioned in Section \ref{sec:prototypical}, at zero temperature, three discontinuities in the second derivative were observable: at $\omega = v_FQ$, at $\omega = v_FQ + 2\mu$ and another branch at $\omega = v_FQ-2\mu$, the latter of which may contribute to plasmon excitations \cite{pines2018}. At finite temperatures, the branches associated with the chemical potential of the Weyl mode remain visible at low temperatures, but are quickly smeared out as the temperature increases as observed in Fig. \ref{fig:7}. In contrast, the contribution from the Weyl fermion collective excitation persists as the discontinuity along the line $\omega = v_FQ$ lingers even at higher temperatures. This highlights the suitability of these signals to be measured in both time-of-flight and triple-axis neutron spectroscopies within experimental conditions.

\section{Discussion}\label{sec:discussion} 
The previous sections illustrate how the nontrivial neutron response of TNMs is specific to these types of exotic materials and can be used to probe for topological phenomena. Remarkably, calculations of the dynamical structure factor for different models of TNMs unveil how the neutron scattering response displays non-analytical behavior arising from the existence of topological nodes in momentum space in addition to unique electromagnetic responses such as the chiral anomaly in the case of WSMs.

In TNMs, the non-analytical smoothness of the neutron scattering signal with respect to wavevector or energy lies along a line $\omega=v_FQ$ which persists even at high temperatures, demonstrating topological robustness as well as a pointer of how one can use this feature to explore the topological nodes in momentum space and determine $v_F$. Furthermore, this characteristic non-analytic smoothness may also exist along lines $\omega = v_FQ \pm 2\mu$ for finite $\mu$ so long as the temperature remains sufficiently low as this signal diminishes with temperature increase. This can be used to experimentally determine parameters associated with the chiral anomaly in WSMs provided that the temperature is sufficiently low.

\begin{figure}[ht!]
	\centering
	\includegraphics[width=\linewidth]{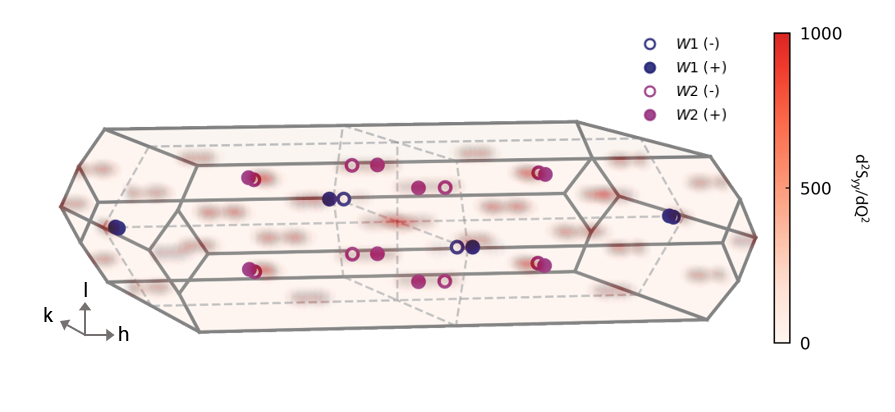}
	\caption{\textbf{Calculation for type-I WSM TaP.} Intensity map of the second derivative of the dynamical structure factor $S_{yy}$ with respect to wavevector $\boldsymbol{Q}$ (averaged along all directions in $Q$-space) in the Brillouin zone of type-I WSM tantalum phosphide (TaP). Elevated intensity is seen at locations corresponding to combinations of nesting wavevectors amongst the different locations of Weyl nodes, the latter illustrated as circles in the plot with indication of subset (blue or magenta) and chirality (open or closed). The calculation shown here does not consider for the chemical potential difference between two Weyl nodes which would lead to higher intensities at $\omega = v_FQ \pm 2\mu$.}
	\label{fig:8}
\end{figure}

As a proof of concept, we apply the calculations made in the previous sections for the case of type-I WSM tantalum phosphide (TaP). TaP crystallizes in a body-centered tetragonal lattice with space group I4$_1$md (109) and point group $C_{4\nu}$ with $a = b = 3.32$\si{\angstrom} and $c = 11.34$\si{\angstrom} and lacks inversion symmetry \cite{xu2016}. Fig. \ref{fig:8} illustrates the calculation of the second derivative of $S_{yy}$ with respect to wavevector $\boldsymbol{Q}$, averaged along all directions of $Q$, within the Brillouin zone of TaP. These calculations reveal a large intensity of this signal in regions that correspond to combinations of nesting wavevectors between the locations of the Weyl nodes (corresponding to divergences at $\omega=v_FQ$). Elevated intensities can also be coincidentally seen near the locations of the Weyl nodes themselves due to the crystal symmetry. For simplicity, the calculation does not account for the chemical potential difference between the W1 and W2 subset of nodes in this material which would further reveal elevated intensities at $\omega=v_FQ \pm 2\mu$.

By performing inelastic neutron scattering with a triple-axis or a time-of-flight experiment and calculating the second derivative of the measured signal, one can extract the nesting wavevectors between the Weyl nodes by examining the location of these elevated intensity points in momentum space and reconstruct the locations of the nodes themselves. This could effectively provide experimental parameters of the Weyl nodes such as momentum space locations and chemical potential.

The calculations performed in this study have mainly focused on isotropic systems. However, several DSMs and WSMs have been shown to exhibit anisotropic topological points. This anisotropy in the response function can be captured in our calculations of $S_{yy}$ by performing a similar substitution for the wavevector $Q^\prime\rightarrow Q$ as described in Ref. \cite{thakur2018} which depends on the anisotropic velocities $\{v_x, v_y, v_z\}$ of the Dirac or Weyl cone. As the magnetic neutron scattering signal depends on the component of the current taken to be in the $\hat{x}$ direction in the picture described in Section \ref{sec:prototypical}, one can orient the sample such that the largest velocity component points in this direction to obtain an increased signal. Conversely, one can also map out the magnitudes of the Fermi velocity components using different orientations of the sample if this information is not known a priori.

This study highlights the use of neutron scattering for probing exotic topological phenomena. In fact, the advantageous opportunity of concurrently applying external magnetic and electric fields during the measurement, which is not applicable with the aforementioned techniques and has possibly enhanced resolution capabilities, positions inelastic neutron scattering as a possible enabler of further discoveries in the realm of TNMs. The direct probing of electronic bands with neutron scattering also differs from probing topology with its lattice degrees of freedom \cite{miao2018, han2020, nguyen2020}.

As the repertoire of TNMs continually enlarges, the requirements of ultrahigh purity and large sample mass for neutron scattering experiments are attainable through gradually-improved synthesis and thus, many materials may be ripe for magnetic field-based neutron scattering studies. These include DSMs, NLSMs, inversion-symmetry breaking WSMs such as the TaAs family and more intriguingly, magnetic WSMs such as Co$_3$Sn$_2$S$_2$ \cite{morali2019,liu2019}, Co$_2$MnGa \cite{belopolski2019}, CeAlGe \cite{suzuki2019,puphal2020} whereby field-driven topological phase transitions may be uncovered. Preliminary works on inelastic neutron scattering serving as a platform for studying exotic phenomena in TNMs have only recently emerged. These include a theoretical overview on probing bulk excitations in type-I WSMs with unpolarized and polarized neutrons \cite{bjerngaard2020} and an experimental study of coupling between Dirac fermions and spin waves \cite{sapkota2020}. A non-exhaustive compilation of other neutron scattering experiments on topological semimetals (including candidates) up until now can be noted in Refs. \cite{park2018, tao2019, zhang2019(2), soh2019, sapkota2020, nguyen2020, han2020, liu2020, cai2020, puphal2020, zhang2021, sukhanov2020, xie2021}. Although these earlier surveys are encouraging, a great deal of work remains to accentuate the potential usefulness of neutron scattering towards characterizing future TNMs sensitive to topological bands.

\section{Conclusion}\label{sec:conclusion}

In summary, we provide a theoretical framework showing that TNMs can have large magnetic neutron scattering response even for non-magnetic materials. The calculation of the dynamical structure factor demonstrates that this large response is uncovered through a non-analytical behavior of the measured spectra, seen as a divergent behavior of the second derivative with respect to wavevector $\boldsymbol{Q}$ or frequency $\omega$. Notably, the high-intensity peaks of the second derivative are sensitive to the location of the Weyl nodes and the chemical potential of the Weyl nodes through the values of $\boldsymbol{Q}$ or $\omega$ at which the divergence takes place. The elevated sensitivity of this signal to these criteria may serve as a valuable method to extract information on parameters that are used to explicitly characterize TNMs, especially WSMs, using an experimental technique that has been hitherto underutilized for this purpose.

Notably, this study showcases that one can perform this type of measurement with triple-axis or time-of-flight neutron spectroscopies to determine the reciprocal-space location of the topological nodes, the magnitude of their Fermi velocities and the characteristic scattering time within these types of materials. In addition, upon application of parallel electric and magnetic fields, one can probe for a unique experimental signature of the chiral anomaly, a defining element of WSMs. Our study opens up new opportunities for the use of neutron scattering as an innovative experimental probe of the nontrivial topology of TNMs and complements the current repertoire of experimental techniques that serve in ongoing characterization efforts of these exotic quantum materials.

\section*{Acknowledgments}
T.N., R.P.P., T.T., N.A., and M.L. acknowledge the support from the U.S. DOE, Office of Science, Basic Energy Sciences, award No. DE-SC0020148. R.P.P. acknowledges the support from FEMSA and the Instituto Tecnologico de Estudios Superiores de Monterrey. N.A. acknowledges the support of the National Science Foundation Graduate Research Fellowship Program under Grant No. 1122374. T.N. and T.T acknowledge support from the Mathworks Fellowship.

\appendix

\section{Magnetic scattering cross section}\label{ap:A} We evaluate the magnetic scattering double differential cross-section term in Eq. \eqref{eq:ddcrossection}. Let $\mathcal{V}$ represent the interaction potential operator of a neutron with the magnetic field. The double differential cross-section is \cite{lovesey1984}
\begin{equation}
\begin{split}
    \label{eqa:ddcrosssection}
    \frac{d^2\sigma}{d\Omega dE_f} = & \sum_{\sigma_i\sigma_f\xi_i\xi_f}p_{\sigma_i}p_{\xi_i}\left|\left\langle\xi_f\boldsymbol{k}_f\sigma_f|\mathcal{V}|\xi_i\boldsymbol{k}_i\sigma_i\right\rangle\right|^2\times\\
    &\left(\frac{m_nV}{2\pi\hbar^2}\right)^2\frac{k_f}{k_i}\delta(\hbar\omega+E_i-E_f)
\end{split}
\end{equation}
\noindent where $\left|\xi\boldsymbol{k}\sigma\right\rangle = \left|\xi\right\rangle \otimes \left|\boldsymbol{k}\right\rangle \otimes \left|\sigma\right\rangle$ is the tensor product of the material state $\left|\xi\right\rangle$, the neutron state $\left|\boldsymbol{k}\right\rangle$, and the neutron spin state $\left|\sigma\right\rangle$. In Eq. \eqref{eqa:ddcrosssection}, $m_n$ is the neutron mass, $V$ is the volume, $\hbar$ is the reduced Planck constant, $E$ is the energy, $p$ is the probability of the neutron being in the spin up or spin down state and the subscripts $i, f$ indicate the initial and final states, respectively. The interaction operator is given by
\begin{equation}
    \label{eqa:interaction}
    \mathcal{V}(\boldsymbol{r}) = -\boldsymbol{\mu_n}(\boldsymbol{r})\cdot\boldsymbol{B}(\boldsymbol{r}),
\end{equation}
where $\boldsymbol{\mu_n}(\boldsymbol{r}) = -\gamma\mu_N\boldsymbol{\sigma_N}$ is the magnetic moment operator of the neutron and $\boldsymbol{B}(\boldsymbol{r})$ is the magnetic field experienced by the neutron. In the definition of $\boldsymbol{\mu_n}(\boldsymbol{r})$, $\gamma$ is the neutron gyromagnetic ratio, $\mu_N$ is the nuclear magneton and $\boldsymbol{\sigma_N}$ is the vector of neutron Pauli matrices. The magnetic field is
\begin{equation}
    \label{eqa:magneticfield}
    \boldsymbol{B}(\boldsymbol{r}) = \frac{1}{c}\int d^3\boldsymbol{r}^{\prime}\ \frac{\boldsymbol{J}(\boldsymbol{r}^{\prime})\times(\boldsymbol{r}-\boldsymbol{r}^{\prime})}{|\boldsymbol{r}-\boldsymbol{r}^{\prime}|^3}
\end{equation}
\noindent where $\boldsymbol{J}$ is the current density operator. We proceed to calculate the matrix element $\left\langle\boldsymbol{k}_f\sigma_f|\mathcal{V}|\boldsymbol{k}_i\sigma_i\right\rangle$ in Eq. \eqref{eqa:ddcrosssection} using Eqs. \eqref{eqa:interaction} and \eqref{eqa:magneticfield}:
\begin{equation*}
\begin{split}
    \label{eqa:matrixelement}
    \langle\boldsymbol{k}_f&\sigma_f|\mathcal{V}|\boldsymbol{k}_i\sigma_i\rangle = \frac{1}{V}\int d^3\boldsymbol{r}\ \left\langle\sigma_f|e^{i\boldsymbol{Q}\cdot\boldsymbol{r}}\mathcal{V}(\boldsymbol{r})|\sigma_i\right\rangle\\
    &= -\frac{1}{V}\int d^3\boldsymbol{r}\ \left\langle\sigma_f\left|e^{i\boldsymbol{Q}\cdot\boldsymbol{r}}\boldsymbol{\mu_n}(\boldsymbol{r})\cdot\boldsymbol{B}(\boldsymbol{r})\right|\sigma_i\right\rangle\\
    &= \frac{4\pi i\gamma\mu_N}{cV}\left\langle\sigma_f\left|\boldsymbol{\sigma_N}\cdot\int d^3{\boldsymbol{r}^{\prime}}\ e^{i\boldsymbol{Q}\cdot\boldsymbol{r}^{\prime}}\frac{\boldsymbol{J}(\boldsymbol{r}^{\prime})\times\boldsymbol{Q}}{Q^2}\right|\sigma_i\right\rangle\\
    &= -\frac{4\pi i\gamma\mu_N}{cV}\left\langle\sigma_f\left|\boldsymbol{\sigma_N}\cdot\frac{\boldsymbol{J}(\boldsymbol{Q})\times\boldsymbol{Q}}{Q^2}\right|\sigma_i\right\rangle
\end{split}
\end{equation*}
\noindent where $\boldsymbol{Q}\equiv\boldsymbol{k}_i-\boldsymbol{k}_f$ is the neutron momentum transfer and $Q = |\boldsymbol{Q}|$. In the derivation above, we defined the Fourier transform of the current operator as
\begin{equation*}
    \label{eqa:ftcurrent}
    \boldsymbol{J}(\boldsymbol{Q})=\int d^3\boldsymbol{r}^{\prime}\ \boldsymbol{J}(\boldsymbol{r}^{\prime})e^{i\boldsymbol{Q}\cdot\boldsymbol{r}^{\prime}}.
\end{equation*}

To proceed further, we define the neutron magnetization operator in momentum space as
\begin{equation}
\begin{split}
    \label{eqa:magnetizationmomentum}
    \boldsymbol{M}(\boldsymbol{Q}) = -i\frac{m_e}{e\hbar}\frac{\boldsymbol{J}(\boldsymbol{Q})\times\boldsymbol{Q}}{Q^2}= -\frac{i}{2\mu_B c}\frac{\boldsymbol{J}(\boldsymbol{Q})\times\boldsymbol{Q}}{Q^2}
\end{split}
\end{equation}
\noindent where $\mu_B=e\hbar/2m_ec$ is the Bohr magneton in Gaussian units. The term in Eq. \eqref{eqa:ddcrosssection} can be simplified as 
\begin{equation}
\begin{split}
    \label{eqa:magnitudematrixelement}
    &\sum_{\sigma_i\sigma_f\xi_i\xi_f}p_{\sigma_i}p_{\xi_i}\left|\left\langle\xi_f\boldsymbol{k}_f\sigma_f|V|\xi_i\boldsymbol{k}_i\sigma_i\right\rangle\right|^2\bigg(\frac{m_nV}{2\pi\hbar}\bigg)^2\approx\\
    &(\gamma r_e)^2\sum_{\xi_i\xi_fa}p_{\xi_i}\left\langle\xi_i|M_a(-\boldsymbol{Q})|\xi_f\right\rangle\left\langle\xi_f|M_a(\boldsymbol{Q})|\xi_i\right\rangle
\end{split}
\end{equation}
where we simplified the prefactor as
\begin{equation*}
    \label{eqa:prefactor}
    \left(\frac{m_nV}{2\pi\hbar^2}\right)^2\left(\frac{8\pi\gamma\mu_N\mu_B}{V}\right)^2=\left(\gamma r_e\right)^2
\end{equation*}
where $r_e$ is the classical electron radius. In Eq. \eqref{eqa:magnitudematrixelement}, we label $a=1,2,3$ for the spatial direction and the correlation is taken with respect to the initial material state. We performed the summation over the neutron Pauli matrices with an assumption of non-polarized neutrons. 

Furthermore, we notice that the energy conservation factor in Eq. \eqref{eqa:ddcrosssection} can be rewritten using the integral representation of the $\delta$-function as
\begin{equation}
\begin{split}
    \label{eqa:energyconservation}
    \sum_{\xi_i\xi_fa}& p_{\xi_i}\left\langle \xi_i\left|M_a(-\boldsymbol{Q})\right|\xi_f\right\rangle\left\langle\xi_f\left|M_a(\boldsymbol{Q})\right|\xi_i\right\rangle \mathcal{I}\\
    &=\int_{-\infty}^\infty dt\ \frac{e^{-i\omega t}}{2\pi\hbar}\left\langle\boldsymbol{M}(-\boldsymbol{Q},0)\cdot\boldsymbol{M}(\boldsymbol{Q},t)\right\rangle
\end{split}
\end{equation}
\noindent where
\begin{equation*}
    \mathcal{I}=\delta(\hbar\omega+E_i-E_f)=\int_{-\infty}^\infty dt\ \frac{e^{-i(\hbar\omega+E_i-E_f)t/\hbar}}{2\pi\hbar}
\end{equation*}
\noindent and $M_a(\boldsymbol{Q},t)=e^{iHt/\hbar}M_a(\boldsymbol{Q},0)e^{-iHt/\hbar}$ is the neutron magnetization operator in the Heisenberg picture. 

The neutron magnetization operator $\boldsymbol{M}$ defined in Eq. \eqref{eqa:energyconservation} is written in a way that is perpendicular to $\boldsymbol{Q}$. In the general case where localized spins may exist, we can rewrite the correlation in Eq. \eqref{eqa:energyconservation} using the full neutron magnetization operator as
\begin{equation*}
    \label{eqa:fullneutronmagneticoperator}
    \boldsymbol{M}(-\boldsymbol{Q})\cdot\boldsymbol{M}(\boldsymbol{Q})=\sum_{jl}\left(\delta_{jl}-\frac{Q_jQ_l}{Q^2}\right)M_j(-\boldsymbol{Q})M_l(\boldsymbol{Q})
\end{equation*}
\noindent where we suppress the time dependence in the notation. Using the correlation of the neutron magnetization operator defined above, we obtain the corresponding magnetic dynamical structure factor $S_{jl}(\boldsymbol{Q},\omega)$ written in Eq. \eqref{eq:magneticdynamicstructurefactor}. 

In a state of thermal equilibrium, the double differential cross-section from Eq. \eqref{eqa:ddcrosssection} can be written in terms of $S_{jl}(\boldsymbol{Q},\omega)$ as
\begin{equation*}
    \label{eqa:ddcrosssection1}
    \frac{d^2\sigma}{d\Omega d\omega}=\frac{k_f}{k_i}(\gamma r_e)^2\sum_{jl}\left(\delta_{jl}-\frac{Q_jQ_l}{Q^2}\right)S_{jl}(\boldsymbol{Q},\omega)
\end{equation*}
\noindent where we used $dE_f=\hbar\ d\omega$ in the derivation.

\section{Magnetic susceptibility}\label{ap:B} We can rewrite the neutron magnetization correlation $M_j(-\boldsymbol{Q})M_l(\boldsymbol{Q})$  using Eq. \eqref{eqa:magnetizationmomentum} as
\begin{equation}
\begin{split}
    \label{eqa:rewriteM}
    &M_j(-\boldsymbol{Q})M_l(\boldsymbol{Q})\\
    &=\left(\frac{1}{2\mu_Bc}\right)^2\left(\frac{\boldsymbol{J}(-\boldsymbol{Q})\times\boldsymbol{Q}}{Q^2}\right)_j\left(\frac{\boldsymbol{J}(\boldsymbol{Q})\times\boldsymbol{Q}}{Q^2}\right)_l \\
    &= \left(\frac{1}{2\mu_Bc}\right)^2\sum_{ikmn}\epsilon_{jik}\epsilon_{lmn}\frac{Q_kQ_n}{Q^4}J_i(-\boldsymbol{Q})J_m(\boldsymbol{Q})
\end{split}
\end{equation}
\noindent where $j, l$ label the spatial direction. Taking the sum over initial and final states and using the same integral representation of the $\delta$-function for energy conservation as in Eq. \eqref{eqa:energyconservation}, we obtain
\begin{equation}
\begin{split}
    \label{eqa:currentcurrent}
    \sum_{\xi_i\xi_f} &p_{\xi_i}\left\langle \xi_i\left|J_i(-\boldsymbol{Q})\right|\xi_f\right\rangle\left\langle \xi_f\left|J_m(\boldsymbol{Q})\right|\xi_i\right\rangle\mathcal{I} \\
    &= \int_{-\infty}^\infty dt\ \frac{e^{-i\omega t}}{2\pi\hbar}\left\langle J_i(-\boldsymbol{Q},0)J_m(\boldsymbol{Q},t)\right\rangle.
\end{split}
\end{equation}

Now we use the fluctuation-dissipation theorem \cite{tauber2014} in order to relate the dynamical structure factor $S_{jl}(\boldsymbol{Q},\omega)$ with the dynamical response function $\Pi_{im}(\boldsymbol{Q}, \omega)$ as
\begin{equation}
    \label{eqa:flucdissi}
    \text{Im}\left[\Pi_{im}(\boldsymbol{Q}, \omega)\right] = -\frac{\pi}{\hbar}\left(1-e^{-\beta\hbar\omega}\right)S_{im}(\boldsymbol{Q},\omega).
\end{equation}
We obtain Eq. \eqref{eq:mdsf} by combining Eqs. \eqref{eqa:rewriteM}-\eqref{eqa:flucdissi}.

\section{Response function of NLSM}\label{ap:C}
We start with the Hamiltonian $H=\sum d_i\sigma_i$, where $d_i(\boldsymbol{k})$ are defined according to Eq. \eqref{eq:d_components}. We rewrite this Hamiltonian in second quantized form and assume $\boldsymbol{k}^2 \rightarrow \psi^\dagger(\hat{\boldsymbol{k}}^2\psi)$. Within this section, we assume $\hbar=e=1$ and restore these factors in the main text. The resulting currents $J_i$ are written as
\begin{equation}
\begin{split}
    \label{eqa:currents}
     J_{x,y} (\boldsymbol{Q})  &= \frac{1}{m} \int \frac{d^3\boldsymbol{k}}{(2\pi)^3}\ \psi^{\dagger}_{\boldsymbol{k}} \left[k_{x,y} + \frac{Q_{x,y}}{2}  \right] \psi_{\boldsymbol{k}+\boldsymbol{Q}}\\
    J_z (\boldsymbol{Q}) &= v_F \int \frac{d^3\boldsymbol{k}}{(2\pi)^3}\ \psi^{\dagger}_{\boldsymbol{k}} \sigma_y \psi_{\boldsymbol{k}+\boldsymbol{Q}}
\end{split}
\end{equation}
where $m$ is the mass, $v_F$ is the Fermi velocity, $k_i$ are components of the momentum, $Q_i$ are components of the neutron wavevector and $\psi_\pm$ are the eigenvectors of $H$
\begin{equation*}
    \psi_{\pm}(\boldsymbol{k}) = \frac{1}{\sqrt{2}}
    \left(
    \begin{array}{c}
    1 \\   \pm \frac{d_1(\boldsymbol{k}) + i d_2(\boldsymbol{k})}{d(\boldsymbol{k})}
    \end{array}
    \right)
\end{equation*}
with eigenvalues $\pm d(\boldsymbol{k}) = \pm\sqrt{\sum d_i^2(\boldsymbol{k})}$. The corresponding Matsubara Green's function $G(\boldsymbol{k},i\omega_n)$ is given by Eq. \eqref{eq:nodallineGreens}. As in the case for Weyl semimetals, we are interested in computing the response function which is given by the current-current correlator
\begin{equation}
\begin{split}
    \label{eqa:nlsmcurrent-current}
    \Pi_{im}(\boldsymbol{Q}, \omega) &= \int d\tau\ e^{-i \omega \tau}\Pi_{im} (\boldsymbol{Q}, \tau)\\
    &= - \langle J_i(-\boldsymbol{Q}) J_m (\boldsymbol{Q})  \rangle
\end{split}
\end{equation}

By substituting the expressions for the currents, shown in Eq. \eqref{eqa:currents}, into the current-current correlator, Eq. \eqref{eqa:nlsmcurrent-current}, we obtain the expression in Eq. \eqref{eq:responsefunctionnlsm} of the main text for the response function in terms of the Green's function. To proceed, we can rewrite the Green's function as
\begin{equation*}
\begin{split}
    \label{eqa:nodallineGreensS}
    G(\boldsymbol{k}, i\omega_n) = \frac{1}{i\omega_n-d(\boldsymbol{k})}S^{(+)}(\boldsymbol{k}) + \frac{1}{i\omega_n+d(\boldsymbol{k})}S^{(-)}(\boldsymbol{k})
\end{split}
\end{equation*}
where we define the following $S^{(\pm)}(\boldsymbol{k})$ matrices

\begin{equation}
\begin{split}
\label{eqa:smatrices}
    S^{(+)}(\boldsymbol{k}) &= 
\frac{1}{2}
\left(
\begin{array}{cc}
1   &  \frac{d_1(\boldsymbol{k})  - id_2(\boldsymbol{k})}{d(\boldsymbol{k})}
\\
 \frac{d_1(\boldsymbol{k})  + id_2(\boldsymbol{k})}{d(\boldsymbol{k})}  &  1
\end{array}
\right)
\\
S^{(-)}(\boldsymbol{k}) &= 
\frac{1}{2}
\left(
\begin{array}{cc}
1   &  -\frac{d_1(\boldsymbol{k})  - id_2(\boldsymbol{k})}{d(\boldsymbol{k})}
\\
 -\frac{d_1(\boldsymbol{k})  + id_2(\boldsymbol{k})}{d(\boldsymbol{k})}  &  1
\end{array}
\right).
\end{split}
\end{equation}

In the following, we use the shorthand notation $v_{i,\boldsymbol{k}} = v_i\left(\boldsymbol{k}\right)$ corresponding to the velocity operator where $v_{x,y} = (1/m)(k_{x,y} + Q_{x,y}/2)$ and $v_z = v_F\sigma_y$. For additional conciseness, we will write $d_{i,\boldsymbol{k}} = d_i(\boldsymbol{k})$, $d_{\boldsymbol{k}} = d(\boldsymbol{k})$ and $S_{\boldsymbol{k}}^{(\pm)} = S^{(\pm)}(\boldsymbol{k})$ when appropriate.

After rewriting in terms of $S^{(\pm)}(\boldsymbol{k})$ matrices, we can insert the Green's function into the polarization operator of Eq. \eqref{eq:responsefunctionnlsm} and perform the summation over Matsubara frequencies to obtain
\begin{widetext}
    \begin{equation*}
    \begin{split}
        \Pi_{im} (\boldsymbol{Q}, \omega) = 
        -  \int \frac{d^3\boldsymbol{k}}{(2\pi)^3}&\Big(
    \frac{n_F(d_{\boldsymbol{k}}) - n_F(d_{\boldsymbol{k+Q}})}{i\omega + d_{\boldsymbol{k}} - d_{\boldsymbol{k+Q}}}
    \mathrm{Tr}\big[v_{i, \boldsymbol{k}+ \frac{\boldsymbol{Q}}{2}} S^{(+)}_{\boldsymbol{k}} v_{m, \boldsymbol{k}+ \frac{\boldsymbol{Q}}{2}} S^{(+)}_{\boldsymbol{k+Q}}\big]\\ 
        &+\frac{n_F(d_{\boldsymbol{k}}) - n_F(- d_{\boldsymbol{k+Q}})}{i\omega + d_{\boldsymbol{k}} + d_{\boldsymbol{k+Q}}}
    \mathrm{Tr}\big[v_{i, \boldsymbol{k}+ \frac{\boldsymbol{Q}}{2}} S^{(+)}_{\boldsymbol{k}} v_{m, \boldsymbol{k}+ \frac{\boldsymbol{Q}}{2}} S^{(+)}_{\boldsymbol{k+Q}}\big]\\
        &+\frac{n_F(-d_{\boldsymbol{k}}) - n_F(d_{\boldsymbol{k+Q}})}{i\omega - d_{\boldsymbol{k}} - d_{\boldsymbol{k+Q}}}
    \mathrm{Tr}\big[v_{i, \boldsymbol{k}+ \frac{\boldsymbol{Q}}{2}} S^{(+)}_{\boldsymbol{k}} v_{m, \boldsymbol{k}+ \frac{\boldsymbol{Q}}{2}} S^{(+)}_{\boldsymbol{k+Q}}\big]\\
        &+\frac{n_F(- d_{\boldsymbol{k}}) - n_F(-d_{\boldsymbol{k+Q}})}{i\omega + d_{\boldsymbol{k}} + d_{\boldsymbol{k+Q}}}
    \mathrm{Tr}\big[v_{i, \boldsymbol{k}+ \frac{\boldsymbol{Q}}{2}} S^{(+)}_{\boldsymbol{k}} v_{m, \boldsymbol{k}+ \frac{\boldsymbol{Q}}{2}} S^{(+)}_{\boldsymbol{k+Q}}\big]
    \Big)
    \end{split}
    \end{equation*}
where $n_F(x)=(e^{\beta x}+1)^{-1}$ is the Fermi-Dirac distribution function. The imaginary part of $\Pi_{im}(\boldsymbol{Q},\omega)$, which appears in the expression for the neutron dynamical structure factor written in Eq. \eqref{eq:mdsf}, has the form
    \begin{equation}
    \begin{split}
    \label{eqa:fourcontributionstoimaginary}
        \text{Im}\left[\Pi_{im} (\boldsymbol{Q}, \omega)\right] = \pi \int \frac{d^3\boldsymbol{k}}{(2\pi)^3}\ &\Big(\left( n_F(d_{\boldsymbol{k}}) - n_F(d_{\boldsymbol{k+Q}}) \right)\mathrm{Tr}\big[v_{i, \boldsymbol{k}+ \frac{\boldsymbol{Q}}{2}} S^{(+)}_{\boldsymbol{k}} v_{m, \boldsymbol{k}+ \frac{\boldsymbol{Q}}{2}} S^{(+)}_{\boldsymbol{k+Q}}\big]\delta\left(\omega + d_{\boldsymbol{k}} - d_{\boldsymbol{k+Q}}\right)\\ 
        &+\left( n_F(d_{\boldsymbol{k}}) - n_F(- d_{\boldsymbol{k+Q}}) \right)\mathrm{Tr}\big[v_{i, \boldsymbol{k}+ \frac{\boldsymbol{Q}}{2}} S^{(+)}_{\boldsymbol{k}} v_{m, \boldsymbol{k}+ \frac{\boldsymbol{Q}}{2}} S^{(+)}_{\boldsymbol{k+Q}}\big]\delta\left( \omega + d_{\boldsymbol{k}} + d_{\boldsymbol{k+Q}}\right)\\ 
        &+\left( n_F(-d_{\boldsymbol{k}}) - n_F(d_{\boldsymbol{k+Q}}) \right)\mathrm{Tr}\big[v_{i, \boldsymbol{k}+ \frac{\boldsymbol{Q}}{2}} S^{(+)}_{\boldsymbol{k}} v_{m, \boldsymbol{k}+ \frac{\boldsymbol{Q}}{2}} S^{(+)}_{\boldsymbol{k+Q}}\big]\delta\left(\omega - d_{\boldsymbol{k}} - d_{\boldsymbol{k+Q}}\right)\\ 
        &+\left( n_F(- d_{\boldsymbol{k}}) - n_F(-d_{\boldsymbol{k+Q}}) \right)\mathrm{Tr}\big[v_{i, \boldsymbol{k}+ \frac{\boldsymbol{Q}}{2}} S^{(+)}_{\boldsymbol{k}} v_{m, \boldsymbol{k}+ \frac{\boldsymbol{Q}}{2}} S^{(+)}_{\boldsymbol{k+Q}}\big]\delta\left(\omega + d_{\boldsymbol{k}} + d_{\boldsymbol{k+Q}}\right)\Big)
    \end{split}
    \end{equation}
\end{widetext}
where $\delta(x)$ is the delta function. We assume $\omega>0$ without loss of generality and take $\boldsymbol{Q} = Q\hat{\boldsymbol{z}}$. Assuming zero Fermi level and zero temperature, the $xx$-component of the imaginary part of the response function is simplified to Eq. \eqref{eq:nodallineresponsezerofermi} because only one of the four contributions in Eq. \eqref{eqa:fourcontributionstoimaginary} remains nonzero as discussed in the main text.

To eliminate the $\delta$-function, we need to take an integral over $k_z$. In order to do so, we compute the derivative $d(\omega-d_{\boldsymbol{k}}-d_{\boldsymbol{k+Q}})/dk_z = -v_F^2(k_z\omega + d_{\boldsymbol{k}}Q)/(d_{\boldsymbol{k}}d_{\boldsymbol{k+Q}})$. By setting the argument of the $\delta$-function to zero and solving for $k_z$, we find the expressions
\begin{equation}
\begin{split}
\label{eqa:deltafunctionarguments}
    k_z &= - \frac{Q}{2} \pm \frac{\omega}{2v_F} \sqrt{1  -  \frac{(k_{\perp}^2 - p_0^2)^2}{m^2(\omega^2 - v_F^2 Q^2)}}\\
    d_{\boldsymbol{k}} &= \frac{\omega}{2} \mp \frac{v_F Q}{2}
\sqrt{1  -  \frac{(k_{\perp}^2 - p_0^2)^2}{m^2(\omega^2 - v_F^2 Q^2)}}\\
    d_{\boldsymbol{k+Q}} &= \frac{\omega}{2} \pm \frac{v_F Q}{2}
\sqrt{1  -  \frac{(k_{\perp}^2 - p_0^2)^2}{m^2(\omega^2 - v_F^2 Q^2)}}. 
\end{split}
\end{equation}

As we require that $d_{\boldsymbol{k}}, d_{\boldsymbol{k+Q}} > 0$, we can solve these inequalities using Eq. \eqref{eqa:deltafunctionarguments} to obtain the condition

\begin{equation*}
    v_F^2 Q^2 - \omega^2 < v_F^2 Q^2 \frac{(k_{\perp}^2 - p_0^2)^2}{m^2(\omega^2 - v_F^2 Q^2)}
\end{equation*}
which, in turn, implies that $\omega^2 - v_F^2 Q^2 > 0$. By enforcing this condition onto Eq. \eqref{eq:nodallineresponsezerofermi} and performing the integral over $k_z$, we can simplify the expression to
\begin{equation}
\begin{split}
\label{eqa:simplifiedIm}
    \text{Im}&\left[\Pi_{xx}(Q\hat{\boldsymbol{z}},\omega)\right]
= \int dk_{\perp} k_{\perp} \sum\limits_{k_{z, \pm}}\left(n_F(- d_{\boldsymbol{k}})  -  n_F( d_{\boldsymbol{k+Q}})  \right)\\&\times\frac{1}{16\pi v_F} \frac{k_{\perp}^2}{m^2} \sqrt{1 - \frac{(k_{\perp}^2 - p_0^2)^2}{m^2 (\omega^2 - v_F^2 Q_z^2)}}
\end{split}
\end{equation}
where the summation over $k_{z,\pm}$ refers to the two possibilities of sign in the expression for $k_z$ in Eq. \eqref{eqa:deltafunctionarguments}. The limits of integration of Eq. \eqref{eqa:simplifiedIm} for $k_{\perp}$ are determined by the condition that the argument in the square roots of Eq. \eqref{eqa:deltafunctionarguments} be strictly positive which is equivalent to
\begin{equation}
\label{eqa:limitsofint}
    0  < k_{\perp}^2  <  p_0^2 + m \sqrt{\omega^2 - v_F^2 Q_z^2}
\end{equation}

Taking $n_F(-d_{\boldsymbol{k}}) - n_F(d_{\boldsymbol{k+Q}}) =  1$ for the case of zero temperature and integrating over $k_{\perp}$ results in the final expression for $\text{Im}\left[\Pi_{xx}(Q\hat{\boldsymbol{z}},\omega)\right]$ shown in Eq. \eqref{eq:nodallineresponsezerofermizerotemp} of the main text.

\twocolumngrid
\bibliographystyle{apsrev4-1}
\bibliography{references}

\end{document}